Load Forecasting in the Era of Smart Grids:

Opportunities and Advanced Machine Learning Models

by

Aurausp Maneshni

A Thesis Presented in Partial Fulfillment
of the Requirements for the Degree
Master of Science

Approved May 2025 by the
Graduate Supervisory Committee:

Yang Weng, Chair
Ahmed Ewaisha
Deana Delp

ARIZONA STATE UNIVERSITY

May 2025

ABSTRACT

Electric energy is difficult to store, requiring stricter control over its generation, transmission, and distribution. A persistent challenge in power systems is maintaining real-time equilibrium between electricity demand and supply. Oversupply contributes to resource wastage, while undersupply can strain the grid, increase operational costs, and potentially impact service reliability. To maintain grid stability, load forecasting is needed. Accurate load forecasting balances generation and demand by striving to predict future electricity consumption. This thesis examines and evaluates four machine learning frameworks for short term load forecasting, including gradient boosting decision tree methods such as Extreme Gradient Boosting (XGBoost) and Light Gradient Boosting Machine (LightGBM). A hybrid framework is also developed. In addition, two recurrent neural network architectures, Long Short Term Memory (LSTM) networks and Gated Recurrent Units (GRU), are designed and implemented. Pearson Correlation Coefficient is applied to assess the relationships between electricity demand and exogenous variables. The experimental results show that, for the specific dataset and forecasting task in this study, machine learning-based models achieved improved forecasting performance compared to a classical ARIMA baseline.

# ACKNOWLEDGMENTS

The completion of this thesis may not have been possible without the support, guidance, and encouragement of many great individuals. First and foremost, I would like to express my deepest gratitude to my advisor and mentor, Dr. Yang Weng, for his support, guidance, and continuous inspiration throughout this research.

I extend my sincere appreciation to my entire defense committee, Dr. Deana Delp and Dr. Ahmed Ewaisha, for their invaluable feedback and review of this research. Special acknowledgements go to Dr. Vijay Vittal for inspiring my interest in Power Systems and Dr. Stefan Myhajlenko for providing me the opportunity to utilize the Goldwater computing lab, which contributed to the progress and success of this research.

I am grateful to Dr. Stevan Hunter from the University of Maryland for his support and for ameliorating my approach to power circuits. My heartfelt gratitude goes to Dr. Darryl Morell, whose mentorship in electrical engineering amped up my passion for the subject since my undergraduate years. Sincere thanks to Dr. Walter for facilitating many of our Zoom sessions and special thanks to my academic advisor, Lynn Pratte.

I am forever grateful to my parents for their unconditional love, patience, and support, which have been the foundation of my academic journey. Their belief in me has given me the strength to persevere through challenges, and their guidance has shaped both my personal and professional growth.

Last but not least, a big thank you to my colleagues and fellow teaching assistants at the Ira A. Fulton Schools of Engineering for their collaboration and camaraderie throughout this endeavor and beyond.

TABLE OF CONTENTS

LIST OF TABLES

LIST OF FIGURES

| Figure | | Page |
|---|---|---|

PREFACE

With the growing electrification of the economy, a much higher demand for electricity and power generation is needed. In the United States, net electricity generation increased by 18% between 2001 and 2023, reaching approximately 4.42 billion megawatt-hours in 2023, as shown in Fig. 1.1 [1]. Electrical energy is produced through various generation sources such as coal, petroleum, natural gas, nuclear, hydroelectric, and other renewable sources. Once generated, electricity is distributed either according to statistical projections or based on real-time demand from consumers. Being able to accurately predict and forecast this demand, or the needed power on a consistent basis, necessitates the need for load forecasting. The predicted load demand will allow the utility companies to efficiency allocate resources and meet the supply-demand of the consumers. The two main gateways to load forecasting include the usage of classical time series or the usage of machine learning algorithms, which has been proven to yield a much higher accuracy while adapting to nonlinearities.

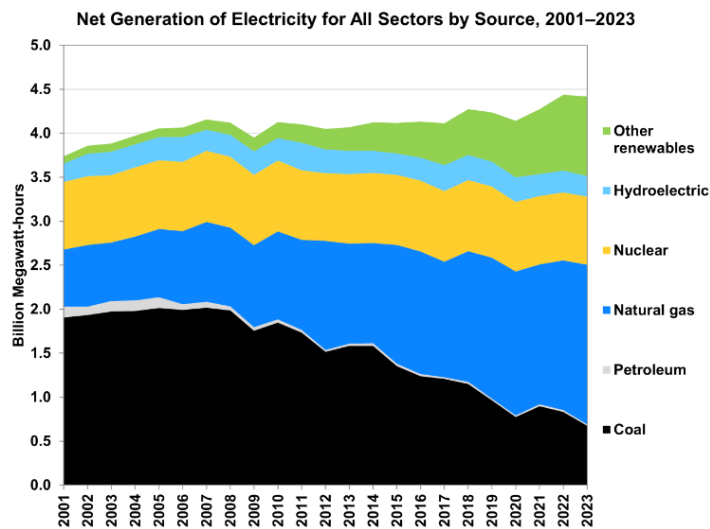

Figure 1.1: U.S. Electrical Energy Sources [1].

CHAPTER 1

INTRODUCTION

*1.1 Background*

Load forecasting is, by concept, a consolidation of methods, implementations, and techniques aimed at predicting future electricity demand (load) based on historical data such as, past consumption usage, weather variations, and consumer to grid behavior. Load forecasting is used in planning, control, and operation of electric power systems. Because electricity cannot be economically stored in large quantities, generation must continuously balance the load to maintain grid stability. Thus, accurate load forecasting enables utility/energy companies and system operators to safeguard sufficient generation in order to meet demand at all times. Moreover, accurate load forecasting helps reduce operational costs by optimizing unit commitment and avoiding over-/under-generation. Accurate load forecasting can further help reduce spinning reserve capacity and therefore schedule device maintenance properly. In the last two sentences, the term *accurate load forecasting* was used. This is because achieving a complete supply-demand equilibrium via forecasting is nearly impossible due to a plethora of uncertainties; these include emergencies, rapid shifts in weather, and other unpredictable variances that can pose unforeseen challenges [2, 3, 4, 5, 6]. However, it worth noting that machine learning enables significantly accurate load forecasting vis-à-vis the challenges.

Machine learning has emerged as a powerful tool capable of realizing complex patterns and nonlinear relations. This is because early electrical load forecasting models were almost always partial to traditional statistical methods. Meaning, a mathematical framework that incorporates statistical assumptions about the load data is generated. Moreover, while machine learning itself is not new technology, advancements in computing power, particularly over the past few years, have enabled significant breakthroughs across various fields, including power and energy [7, 8, 11]. Power system studies have traditionally relied on physical model-driven methods for decades. When it comes to load forecasting, classical time series modeling has traditionally been the go-to approach, relying on pre-determined, already derived statistical methods [9]. However, these models often struggle with nonlinear dependencies and other external influences, thus limiting their accuracy. Conversely, machine learning is making it possible to address constantly changing and nonlinear data without relying on pre-determined models [8]. In short, machine learning has become an integral and evolving method for improving the accuracy of load forecasting in modern power systems.

*1.2 Research Motivation*

This thesis explores the comparative performance of classical time series and machine learning-based load forecasting methods. A baseline model is established using ARIMA as reference. Pearson correlation coefficients are used for feature selection, thereby classifying exogenous variables that impact the load demand behavior. Their importance is visually represented. Moreover, four machine learning models are developed, trained, and implemented. The models selected in this thesis are tested on real-world load demand data and compared against the baseline ARIMA model to evaluate their predictive accuracy. Finally, the results demonstrated in this work prove the advantages of machine learning in power load forecasting.

*1.3 Thesis Outline*

Following this chapter, the remaining chapters of this thesis are organized as follows: Chapter 2 introduces the classical time series load forecasting models and discusses the conventional approaches in load forecasting. Chapter 3 begins by discussing the frameworks of machine learning methodologies for load forecasting. Chapter 3 also discusses some of the challenges associated with load forecasting using machine learning. In Chapter 4, the methodology of this research is outlined, including the treatment of exogenous factors influencing power load, processing techniques such as handling atypical data, gap correction, and normalization. This chapter also describes the metrics used for forecast evaluation and feature selection approach. Chapter 5 presents the implementation of forecasting models, including a hybrid model. It begins with a baseline classical model, followed by the hybrid model, and four machine learning approaches. Chapter 6 discusses the experimental setup and results. It begins with the Pearson correlation analysis results and evaluates the performance of each forecasting model in terms of its prediction accuracy and reliability. Chapter 7 provides a summary of the model accuracy results as well as potential areas for future research in load forecasting. Chapter 8 draws the conclusion.

CHAPTER 2

CLASSICAL MODELS

*2.1 Classification of Load Forecasting Ranges*

Currently, there is no unanimously accepted standard for categorizing load forecast ranges [18]. There are also some differences in definitions of forecast ranges. However, researchers generally classify load forecasting based on prediction duration into four classes: ultra/very short-term load forecasting (VSTLF), short-term load forecasting (STLF), medium-term load forecasting (MTLF), and long-term load forecasting (LTLF) [10, 11, 12]. Fig. 2.1 illustrates the classification of load forecasting ranges based on prediction timeframes.

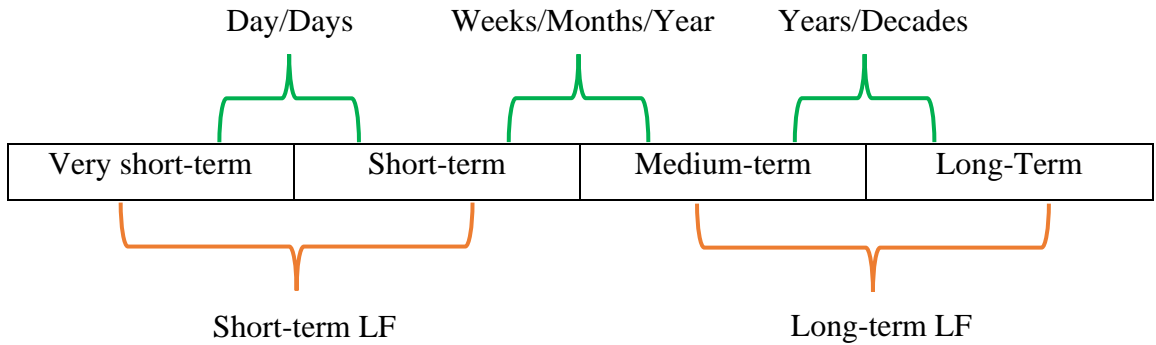

Figure 2.1: Classification of Load Forecasting Ranges Based on Prediction Durations.

Load forecasting is classified based on prediction timeframes, which determine the forecasting horizon. These timeframes influence the selection of appropriate modeling techniques which are discussed in this research. Very short-term load forecasting focuses on predicting power demand within the next few minutes, primarily used for real-time grid operation monitoring. Short-term forecasts typically span from an hour, to one week. This forecast plays an imperative role in the day-to-day operations of utilities such as unit commitment, economic dispatch, and load management. Moreover, short-term load forecasting is commonly referred to as a daily load forecast. Medium-term load forecasting spans over several weeks, to months, and up to one year. This is necessary for fuel procurement, scheduling unit maintenance, energy trading and revenue assessment for the utilities. Finally, long-term load forecasting spans over one year and can range up to 20 years ahead. This type of load forecasting is predominantly used for various infrastructure planning, generation capacity expansion, and also in policy making [10, 11, 12, 13, 14]. Accurately forecasting demand across these timeframes helps maintain the reliability and stability of power systems.

Table 2.1: Classification of Load Forecasting Time Horizons and Applications.

| # | Class | Forecast Horizon | Synopsis of Primary Applications |
|---|-------|------------------|----------------------------------|
| 1 | VSTLF | Minutes $\leq f \leq$ Hours | Real-time grid operation monitoring. |
| 2 | STLF | Hours $\leq f \leq$ Weeks | Unit commitment, economic dispatch. |
| 3 | MTLF | Weeks $\leq f \leq$ 1 Year | Fuel procurement, maintenance scheduling. |
| 4 | LTLF | 1 Year $\leq f \leq$ 20 Years | Infrastructure planning, generation expansion. |

Each forecasting horizon requires different methodologies and data inputs to address the varying levels of uncertainty and influencing factors. Table 2.1 presents a classification of load forecasting time horizons along with their primary applications. As power systems evolve, accurate load forecasting supports resource allocation, infrastructure development, demand response, and commitment planning. With this foundation, the next section explores the conventional approaches in load forecasting. These approaches will be studied and compared with machine learning models used in this research, following the future chapters.

## *2.2 Conventional Approaches in Load Forecasting*

Traditional load forecasting methods have long been used to predict electricity demand based on historical patterns and statistical relationships. These methods primarily rely on time series analysis and regression techniques to model load behavior over different forecast horizons. Among them, the Autoregressive Integrated Moving Average (ARIMA) model is recognized as a conventional approach for short-term and medium-term load forecasting due to its ability to capture temporal dependencies in historical data [9, 15]. ARIMA models, based on the Box-Jenkins method, operate under the assumption that future values can be predicted based on past observations within a time series, and therefore, making them an effective conventional approach in load forecasting [16, 17]. However, despite their effectiveness, these traditional models will struggle with nonlinear relationships, external factors, and sudden fluctuations in electricity demand.

### *2.2.1 Box-Jenkins Basic Models*

The Box-Jenkins is a systematic approach to time series modeling, entailing Autoregressive (AR), Moving Average (MA), Autoregressive Moving Average (ARMA), and Autoregressive Integrated Moving Average (ARIMA) models [16]. The latter model will be discussed in more detail. The following subsections provide an overview of the above models and are as follows:

1. Autoregressive (AR) Model:

The AR model expresses the current value of a time series, $X_t$, as a linear combination of its previous/past loads. Thus, the AR model can be used to forecast future load values. Mathematically, the AR model can be written as:

$$X_t = \phi_1 X_{(t-1)} + \phi_2 X_{(t-2)} + \cdots + \phi_p X_{(t-p)} + \epsilon_t \qquad (2.1)$$

$$X_t = \sum_{i=1}^{p} \phi_i X_{t-i} + \epsilon_t \qquad (2.2)$$

Where $\phi_i$ are the unknown AR coefficients, while $\epsilon_t$ is stochastic white noise component. Order of the model determines how many lagged past values are involved. Hence, the AR model can predict future behavior based on past behaviors. The AR model is particularly useful in load forecasting when there is some correlation between the current values of $X_t$ in a time series and its past values [17, 18].

2. Moving Average (MA) Model:

The MA model represents a moving average process by using a linear regression approach, where the current values are expressed as a function of the white noise from one or more previous steps. Mathematically, the MA model is represented:

$$X_t = \epsilon_t + \theta_1 \epsilon_{(t-1)} + \theta_2 \epsilon_{(t-2)} + \cdots + \theta_q \epsilon_{(t-q)} \qquad (2.3)$$

$$X_t = \epsilon_t + \sum_{j=1}^{q} \theta_j \epsilon_{t-j} \qquad (2.4)$$

Where $\theta_j$ are the moving average coefficients. The MA model captures short-term dependencies by smoothing fluctuations and thus become suitable for modeling times series data with unpredictable variations. Hence, the MA model is effective in with random shocks and reducing the noise in time series forecasting [17, 18].

3. Autoregressive Moving Average (ARMA) Model:

   The ARMA model combines both AR and MA components. It captures dependencies in stationary time series data. The AR component accounts for relationships between past and present values, while the MA component models the influence of past white noise errors [17, 18]. An ARMA(p, q) model is mathematically represented as:

   $$X_t = \sum_{i=1}^{p} \phi_i X_{t-i} + \epsilon_t + \sum_{j=1}^{q} \theta_j \epsilon_{t-j} \qquad (2.5)$$

   Where *p* is the number of autoregressive terms, *q* is the number of moving average terms, $\phi_i$ are AR coefficients, $\theta_j$ are MA coefficients, and $\epsilon_t$ is white noise.

*2.2.2 ARIMA Model*

The AR, MA, and consolidated ARMA models were mathematically represented in Equations 2.1 through 2.5. These models can only be used for stationary time series data. From a load forecasting perspective, the ARMA*(p, q)* model is inadequate for describing non-stationary times series. Therefore, ARIMA models were introduced to include the case of non-stationary. An ARIMA model is characterized by three terms *p, d, q* and thus written as ARIMA*(p, d, q)*; where *p* is the number of AR terms, *d* is the minimum number of differencing operations to make the original time series stationary, and $q$ is the number of MA terms [17]. The formulation of the ARIMA(p, d, q) model using lag polynomials is given below:

$$\phi(B) \cdot \nabla^d y_t = \theta(B) \cdot \epsilon_t \qquad (2.6)$$

$$\left[1 - \sum_{i=1}^{p} \phi_i B^i\right] \cdot (1-B)^d y_t = \left[1 + \sum_{j=1}^{q} \theta_j B^j\right] \cdot \epsilon_t \qquad (2.7)$$

Where $B$ represents the lag operator, also known as backshift operator, and the remaining variables were introduced in Equations 2.1 through 2.5. ARIMA uses linear combinations of past values and errors. This means it cannot inherently capture nonlinear patterns in load (for instance, the sharply increasing evening peak in some regions due to thermostatic load behavior might not be linear). Moreover, ARIMA requires the time series to be stationary, or stationary after differencing, which might not hold true if there are evolving consumption patterns or heteroscedastic variances; any unaddressed non-stationarity can degrade the forecast accuracy [8, 17, 18, 19]. Due to these limitations in ARIMA, this necessitates the need for machine learning-based load forecasting, as it can account for the aforementioned nonlinear traits and patterns.

### 2.2.3 SARIMA Model

The SARIMA model is an extension to the ARIMA model. Its key advantage is its ability to handle seasonality explicitly. With the inclusion of seasonal terms, SARIMA can model the periodic nature of electricity demand better than a standard ARIMA [20]. The general SARIMA equation can be written as:

$$\phi(B)\Phi(B^m)(1-B)^d(1-B^m)^D y_t = \theta(B)\Theta(B^m)\epsilon_t \qquad (2.8)$$

Which can be further expanded as:

$$\left(1 - \sum_{i=1}^{p} \phi_i B^i\right)(1-B)^d y_t = \left(1 + \sum_{j=1}^{q} \theta_j B^j\right)\epsilon_t \quad (2.9)$$

$$\left(1 - \sum_{k=1}^{P} \Phi_k B^{km}\right)(1-B^m)^D y_t = \left(1 + \sum_{\ell=1}^{Q} \Theta_\ell B^{\ell m}\right)\epsilon_t$$

$$\left(1 - \sum_{i=1}^{p} \phi_i B^i - \sum_{k=1}^{P} \Phi_k B^{km}\right)(1-B)^d (1-B^m)^D y_t \quad (2.10)$$

$$= \left(1 + \sum_{j=1}^{q} \theta_j B^j + \sum_{\ell=1}^{Q} \Theta_\ell B^{\ell m}\right)\epsilon_t$$

Where $\sum_{i=1}^{p} \phi_i B^i$ represents the non-seasonal AR terms, $\sum_{j=1}^{q} \theta_j B^j$ represents the non-seasonal MA terms, and $\sum_{k=1}^{P} \Phi_k B^{km}$, $\sum_{\ell=1}^{Q} \Theta_l B^{\ell m}$ represent the seasonal AR and MA terms respectively. While SARIMA improves on ARIMA by addressing seasonality, it still shares many limitations of its non-seasonal counterpart. It remains a linear model, thus, it cannot easily capture nonlinear relationships in the data (such as saturation effects or nonlinear temperature-load relationships). Additionally, it also can become cumbersome if there are multiple seasonal patterns of different lengths or if the seasonal period is not constant. For instance, holidays can disrupt weekly seasonality. SARIMA by itself has no mechanism to handle calendar events or regime changes except by adding indicator variables or treating them as outliers [19, 21, 22]. SARIMA performs well for linear and periodic patterns, but it struggles with nonlinear, irregular, or long-term trend-dependent load behavior. This limitation has contributed to the increasing interest in machine learning techniques for load forecasting. Approaches such as neural networks and gradient boosting have shown improved capability in modeling nonlinear load behavior and incorporating exogenous factors, including weather conditions and demand-side variability [47].

### 2.2.4 Grey Model

Grey model (GM) leverages a small dataset to construct differential equations for short-term load forecasting. Initially, historical data undergoes an accumulated transformation to create a smoother sequence, thus reducing randomness. Then, differential equations are formulated based on the transformed data. The GM(1, 1) model, for instance, uses a first-order equation with one variable. Unlike conventional methods that require large datasets for training, Grey Model performs well with limited data. It is particularly effective for exponentially growing load patterns, but its accuracy decreases when forecasting stationary load curves [23]. As the Grey model assumes exponential growth trend, it becomes less effective for stationary or highly fluctuating load patterns. Additionally, its reliance on small dataset can lead to reduced accuracy compared to data-driven machine learning approaches when large historical data are already available by the utility companies.

### 2.2.5 Multiple Linear Regression Model

Multiple linear regression (MLR) models the relationship between a dependent variable $y$ and multiple independent variables $x_1, x_2, ..., x_k$ using a linear equation. The objective of MLR is to establish a function that captures these relationships, thus enabling predictions based on input features. In load forecasting, factors such as temperature, humidity, and other weather data can influence the demand. The mathematical representation of the MLR model is written in Equation 2.11 below:

$$y = \beta_0 + \beta_1 x_1 + \beta_2 x_2 + \cdots + \beta_k x_k + \epsilon \qquad (2.11)$$

$$y = \beta_0 + \sum_{i=1}^{k} \beta_i x_i + \epsilon \qquad (2.12)$$

Where *y* represents the short-term load to be predicted, $x_i$ are the independent variables affecting power load, and $\beta_i$ are the regression coefficients, $\epsilon$ is the random error term. Just like the previously explained models, MLR also assumes a linear relationship between the predictors and the target variable [24]. As energy consumption patterns become less linear, MLR methods become less effective at load forecasting due to such nonlinearities.

*2.3 Comparative Advantages and Limitations*

Each classical model discussed in this chapter has its strengths and weaknesses in short-term load forecasting. A summary comparison is as follows:

    i.    Time Series forecasting models such as ARIMA and SARIMA are traditionally utilized for capturing regular patterns and autocorrelations in load data. While they have a strong theoretical foundation and offer value as benchmarking tools, their use in actual utility operations is often limited to hybrid setups or vendor-implemented forecasting platforms that integrate multiple techniques, including physics-based or forecast-driven models [46]. Moreover, structural changes, nonlinear effects, and multiple input variables present challenges, as the underlying assumption is that historical patterns repeat in a linear manner [8, 16, 19]. When load is influenced by unpredictable factors or when relationships are non-linear, pure time series models falter. ARIMA and SARIMA also require

parameter tuning (identifying *p, q, P, Q,* etc.), but when properly tuned, the models can serve as reliable baseline forecasters. SARIMA in particular addresses one major limitation of ARIMA by incorporating seasonality [20, 22].

ii. Grey Model: Stands out for minimal data requirements and simplicity. It's very useful in niche cases such as initial load forecasting for a new service area or when data is scarce/unreliable. The Grey model can be quickly computed and can sometimes even beat heavier models in certain comparisons. On the downside, if one does have a lot of data and rich patterns, relying solely on a Grey model is not advisable – it won't utilize the data's full information content [23]. Finally, Grey models need adjustments to handle seasonality; without hybridization, they cannot capture periodic fluctuations.

iii. Multiple Linear Regression (MLR): An approach that directly incorporates key explanatory variables. In load forecasting, its big advantage is integrating weather and calendar effects linearly, which often explains a large portion of demand variation (e.g., temperature vs. load linear relationship during moderate weather). MLR's weaknesses are the assumption of linearity and independence of errors. They might underperform if the true relationship is complex (which it often is for extreme weather impacts on load) [24].

Each classical model discussed above has its advantages and disadvantages in short-term load forecasting. The classification of these forecasting methods is illustrated in Fig. 2.2. To conclude this chapter, classical time series and regression models remain relevant as baseline techniques and as components of hybrid approaches [51]. Thus, while classical models have served and continue to serve well as load forecasting techniques, the evolving demands on forecast accuracy and the availability of big data by the utility companies has spurred a shift toward machine learning-based models that can overcome the limitations of classical approaches, and thus, better handle the challenges of modern power systems load forecasting.

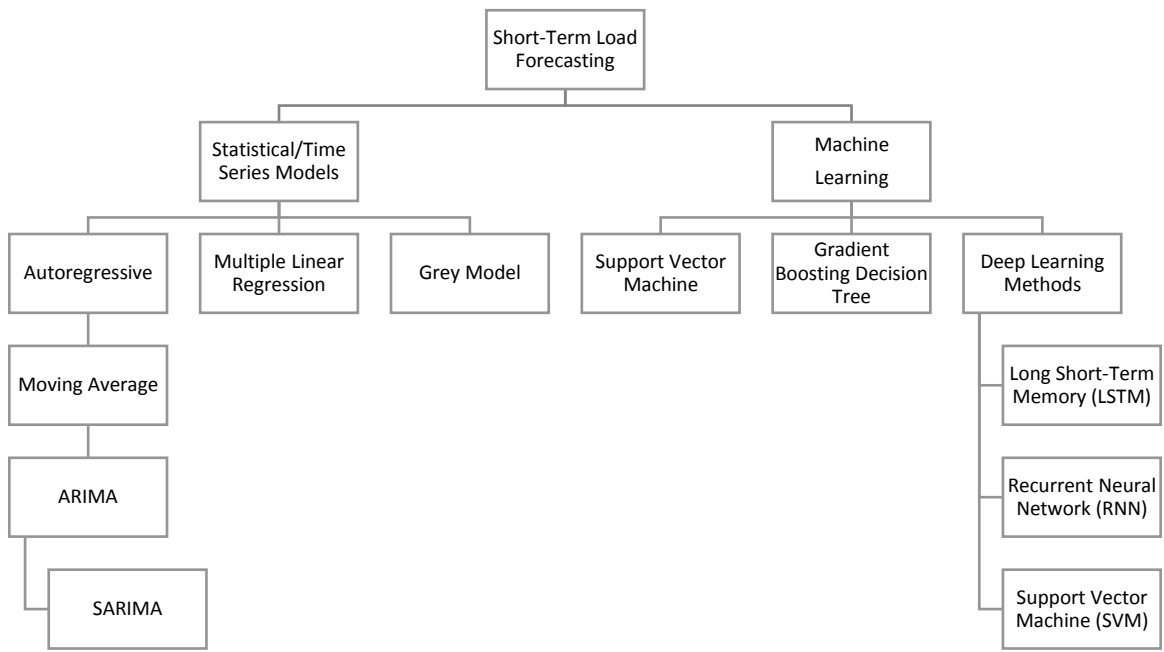

Figure 2.2: Classification of Short-Term Load Forecasting Methods.

CHAPTER 3

MACHINE LEARNING

*3.1 Methods Overview*

Machine learning models have shown improved performance in various short-term load forecasting tasks, especially when modeling nonlinear relationships. However, their advantage depends on data characteristics, model forecast, and the type and availability of exogenous variables. In some scenarios, classical or hybrid models may remain competitive [48]. On this note, while statistical methods such as MLR have been used for short-term load forecasting, ML techniques, such as Support Vector Machines (SVM), Gradient Boosting Machines (GBM), and Deep Learning methods levy greater adaptability to nonlinearities in load data [2, 25]. As noted previously, this thesis aims to examine and analyze ML methods for short-term load forecasting and as such, compare the experimental results with the experimental classical modeling of short-term load forecasting, referenced in the previous chapter. To begin, a top-down approach to ML methodologies are given.

*3.1.1 Support Vector Machine (SVM) & Support Vector Regression (SVR)*

SVM is a supervised learning algorithm that constructs a hyperplane in a high-dimensional space to perform classification. When the data is not linearly separable, SVM uses a kernel function to project the data into a higher-dimensional space, where a linear classifier can be applied. For STLF, SVM is used in its regression form, Support Vector Regression (SVR). For regression, the hyperplane's normal vector includes a function that helps minimize the difference between predicted and actual values [26, 27].

This in return, allows the SVM model to model the data distribution well. Unlike traditional methods, SVM does not rely on prior assumptions about the data; this makes it apt for both stationary and non-stationary load datasets. In STLF, SVR constructs a function $f(x)$ that has at most $\varepsilon$ deviation from the actual target values for all training data, while keeping the model as flat as possible. The regression function takes the form:

$$f(x) = \mathbf{w}^\top \phi(x) + b$$

Where $\phi(\cdot)$ maps the input data to a higher-dimensional feature space, $\mathbf{w}$ represents the weight factor, and $b$ is the bias term. The primal optimization problem for SVR is expressed as:

$$\min_{w,b,\xi_i,\xi_i^*} \frac{1}{2}||w||^2 + C \sum_{i=1}^{n}(\xi_i + \xi_i^*) \qquad (3.2)$$

Subject to:

$$\begin{cases} y_i - \mathbf{w}^\top \phi(x_i) - b \le \varepsilon + \xi_i \\ \mathbf{w}^\top \phi(x_i) + b - y_i \le \varepsilon + \xi_i^* \\ \xi_i, \xi_i^* \ge 0 \end{cases} \qquad (3.3)$$

Here, $\varepsilon$ defines the margin of tolerance, $\xi_i$ and $\xi_i^*$ are slack variables allowing some errors, and $C$ is a regularization parameter controlling the penalty for derivations. The dual formulation, which allows the use of kernel functions $K(x_i, x_j) = \phi(x_i)^\top \phi(x_j)$, is:

$$\begin{aligned} \max_{a_i, a_i^*} &-\frac{1}{2}\sum_{i=1}^{n}\sum_{j=1}^{n}(a_i - a_i^*)(a_j - a_j^*)K(x_i, x_j) \\ &+ \sum_{i=1}^{n}(a_i - a_i^*)y_i - \varepsilon \sum_{i=1}^{n}(a_i - a_i^*) \end{aligned} \qquad (3.4)$$

Subject to:

$$\sum_{i=1}^{n}(a_i - a_i^*) = 0, \ \ 0 \leq a_i, a_i^* \leq C \qquad (3.5)$$

The final regression function used for forecasting is:

$$f(x) = \sum_{i=1}^{n}(a_i - a_i^*) K(x_i, x) + b \qquad (3.6)$$

In STLF, SVR models the relationship between historical load profiles and influencing variables to predict future load demand. A typical feature vector, $x_i$, can include the following:

$$x_i = [Load_{t-1}, Load_{t-2}, \ldots, Temperature, Hour\ of\ Day, Day\ of\ Week, Holiday]$$

To visually illustrate the application of SVR in STLF, a simulated forecasting scenario is presented in Fig 3.1 below using the load profile and corresponding model predictions.

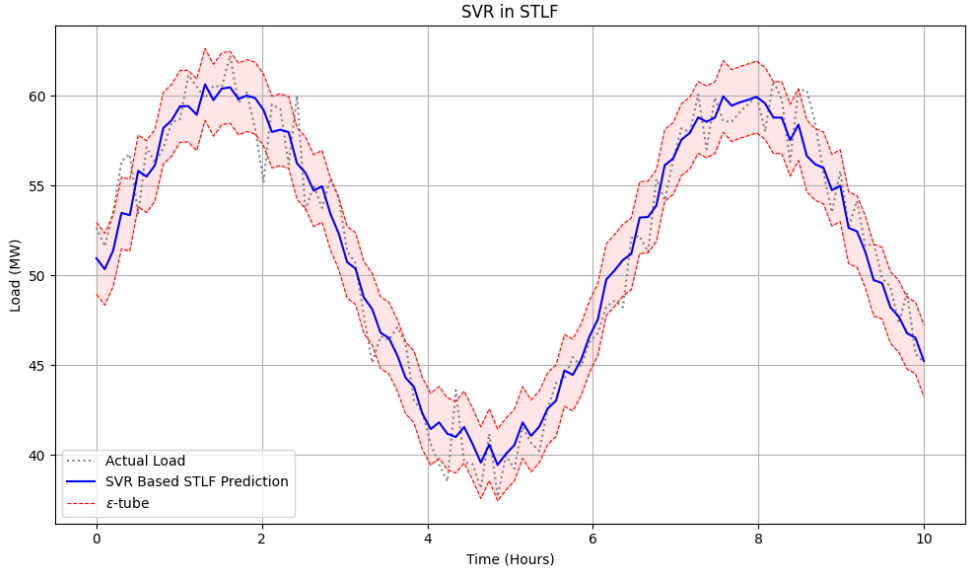

Figure 3.1: SVR Simulation with Ɛ-Insensitive Tube for STLF.

As shown in Fig. 3.1 above, the blue line represents SVR model's predicted load values over time, while the dotted grey line depicts the actual load data. The red dashed lines mark the upper and lower bounds of the Ɛ-insensitive tube, which defines the margin within which prediction errors are not penalized. Lastly, the shaded red region between the boundaries indicates the tolerance band. Following the SVR simulation, Fig. 3.2 simulates how SVR handles nonlinear relationships in load data. Thus, simulating the effect of kernel function in transforming the input feature space. The left plot shows two classes in their original 2-D space, where linear separation is not possible. The right plot displays the same data after kernel mapping into a higher-dimensional space, where the classes become linearly separable. Such transformation permits SVR to recognize and model nonlinear patterns in STLF more accurately.

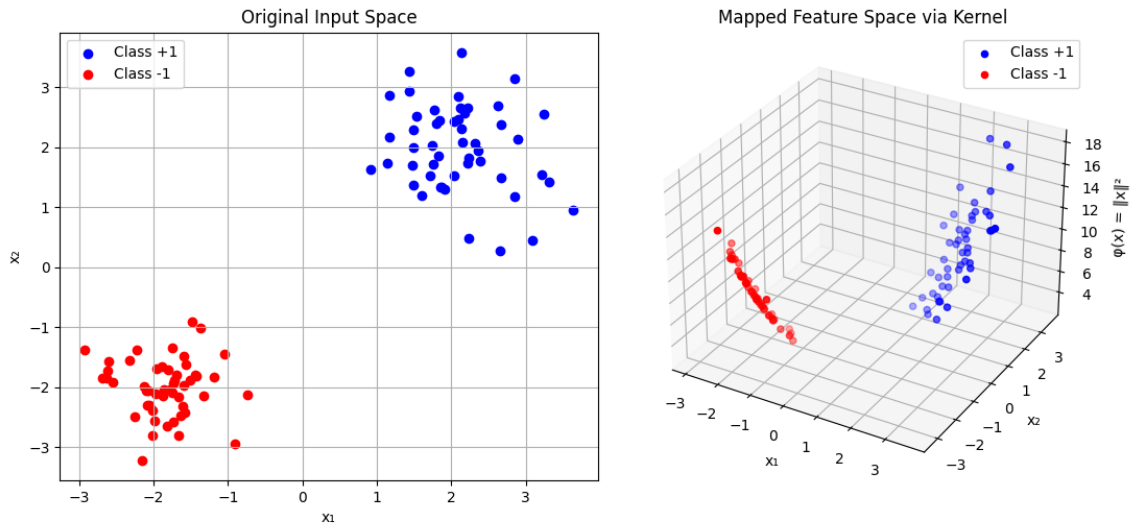

Figure 3.2: Kernel-Based Feature Mapping for Nonlinear Separation in SVR.

### *3.1.2 Artificial Neural Networks*

Artificial Neural Networks (ANNs) are used in load forecasting with decent success due to their ability to model nonlinear, non-stationary relationships between input features and future load values. Compared to SVM, ANNs can model non-stationary load behavior more accurately and be more flexible when dealing with large and nonlinear datasets. ANNs can automatically learn internal feature representations, making them better suited for capturing temporal dependencies in load forecasting tasks. Common architectures of ANNs can include Feed-Forward Neural Network (FFNN) and Recurrent Neural Networks (RNNs) with compositions such as Long Short-Term Memory (LSTM) as an example [28, 29]. A challenge seen when modeling ANNs, particularly with recurrent architectures, is the risk of vanishing or exploding gradients during backpropagation through time, which can hinder the learning of long-term dependencies. Although LSTM networks address this issue to a significant extent by incorporating gating mechanisms, these often require significant computational resources and hyperparameter tuning [30]. Despite these challenges, ANN-based models have shown strong performance in load forecasting and are serving as more mainstream methods of prediction.

The mathematical representation of an ANN for load forecasting involves an input layer, one or more hidden layers, and an output layer. Equation 3.7 defines the forward pass of a single hidden-layer ANN, where the input feature transformed through weighted connections and nonlinear activation functions to produce the predicted load output. Given an input vector $x_i \in R^d$ (i.e., previous loads time, temperature), a single hidden-layer ANN estimates the future load $\hat{y}_i$ as:

$$\hat{y}_i = f\left(\sum_{j=1}^{H} w_j^{(2)} \cdot \sigma\left(\sum_{k=1}^{d} w_{jk}^{(1)} x_{ik} + b_j^{(1)}\right) + b^{(2)}\right) \qquad (3.6)$$

Where $H$ is the number of hidden neurons, $w_{jk}^{(1)}$ weights from input layer to hidden layer, $w_j^{(2)}$ weights from hidden layer to output layer, $b_j^{(1)}$ & $b^{(2)}$ are biases for hidden and output layers, $\sigma(\cdot)$ is the activation function (i.e., ReLU), and finally, $f(\cdot)$ is the output activation (linear for regression). This model can be extended to deep architectures or recurrent variants such as RNN/LSTM for temporal modeling.

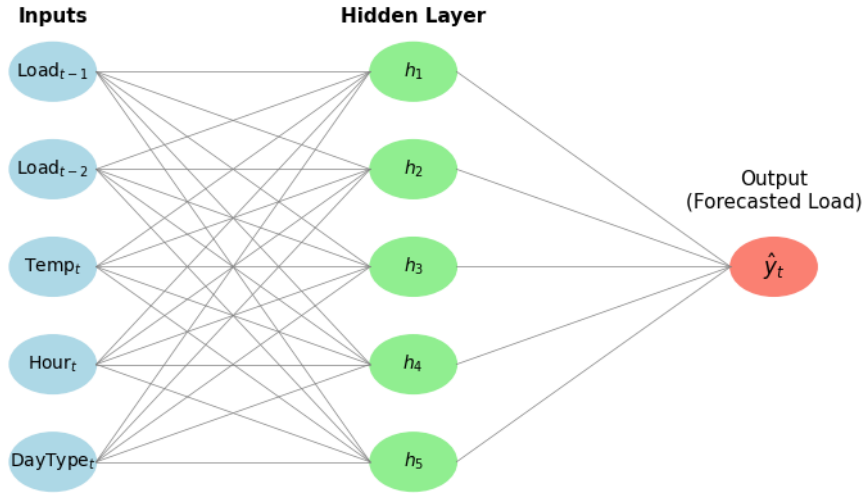

Figure 3.3: Architecture of a Single Hidden-Layer ANN for STLF.

Figure 3.3 illustrates the architecture of a single hidden-layer Artificial Neural Network (ANN) used for short-term load forecasting. The model receives multiple input features, such as previous load values, temperature, hour of the day, and day type, which are used as predictors in forecasting tasks. These inputs are fed into a hidden layer containing of several neurons, each applying a nonlinear activation function. The processed outputs from the hidden layer are then combined and passed to the output layer, which generates the predicted load value $\hat{y}_t$. Finally, this structure captures the relationship between historical and contextual features and the future load demand. The dense connections between layers allow the network to learn the nonlinear dependencies for accurate short-term predictions.

*3.1.3 Gradient Boosting Decision Trees*

Gradient Boosting Decision Trees (GBDT) are ensemble learning techniques that have demonstrated strong performance in load forecasting tasks. The core principle of boosting is to sequentially combine multiple weak learners—typically shallow decision trees—into a single, strong predictive model. Each new tree is trained to correct the residual errors of the previous ensemble using gradient descent optimization on a defined loss function. As the model iteratively learns from its own mistakes, it becomes more accurate over time. This iterative refinement process makes GBDTs particularly effective in capturing complex, nonlinear relationships between historical load values, weather data, calendar features, and other exogenous variables. Unlike traditional statistical models, GBDTs do not require strict assumptions about the data distribution and can automatically handle feature interactions, missing values, and mixed data types. Popular GBDT frameworks such as XGBoost and LightGBM are used in modern load forecasting systems due to their speed, scalability, and dynamism. Moreover, based multiple studies, these models have been effectively applied to real-world power systems for day-ahead, hour-ahead, and even minute-ahead load prediction, often yielding higher accuracy than classical methods such as ARIMA, SARIMA or MLR [31, 32, 33]. Given a training dataset with *n* observations:

$$\{(x_i, y_i)\}_{i=1}^{n} \qquad (3.7)$$

Where $x_i \in R^d$ input feature (i.e., previous loads, temperature, time), $y_i \in R$ actual load at time *t*, GBDT builds an additive model of *M* decision trees:

$$\hat{y}_i = F_M(x_i) = \sum_{m=1}^{M} \gamma_m h_m(x_i) \qquad (3.8)$$

Where $h_m(\cdot)$ is the *m*-th regression tree (weaker learner), $\gamma_m$ is the learning rate or shrinkage factor, and $F_M(x_i)$ is the final predicted load at input $x_i$. Moreover, each tree, $h_m$ is trained to predict the negative gradient (residual) of the loss function $\mathcal{L}$ at step *m*:

$$r_i^{(m)} = -\left[\frac{\partial \mathcal{L}(y_i, F_{m-1}(x_i))}{\partial F_{m-1}(x_i)}\right] \qquad (3.9)$$

A new tree is then fit to the residuals:

$$h_m(x) \approx r^{(m)} \qquad (3.10)$$

and the model is rationalized as:

$$F_m(x) = F_{m-1}(x) + \gamma_m h_m(x) \qquad (3.11)$$

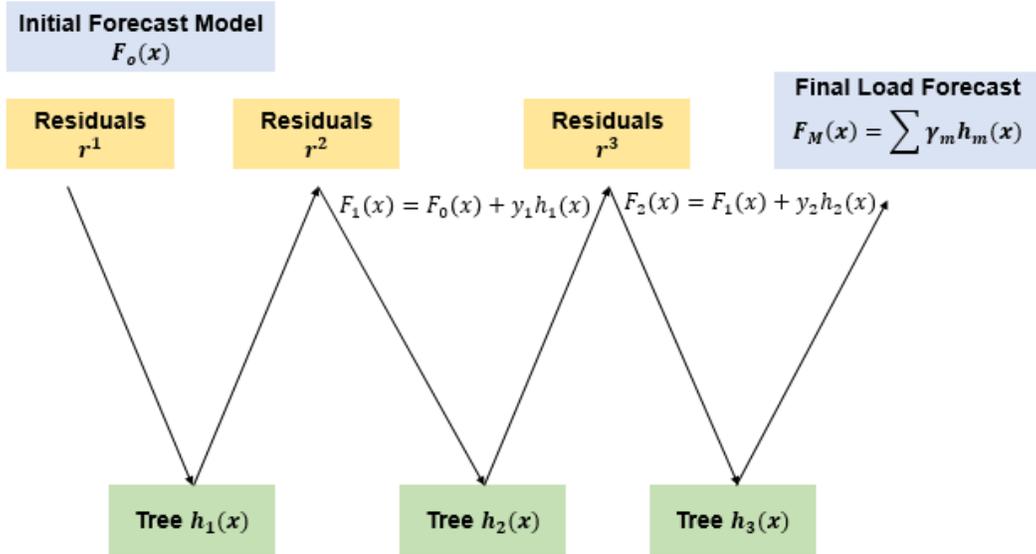

Figure 3.4: Visualization of the Iterative Boosting Process in GBDT, where each learner is trained on residuals and added sequentially to improve load forecasting accuracy.

Figure 3.4 illustrates the GBDT process used in most short to medium-term load forecasting. The process begins with an initial model $F_0(x)$, typically a constant value representing a simple prediction. At each subsequent stage, decision trees $h_1(x), h_2(x),…,$ are sequentially trained to fit the residuals from the previous model. Moreover, these residuals represent the part of the target load not yet captured by the ensemble. The model is then updated iteratively by adding each new tree's prediction, scaled by a learning rate $\gamma$, to the existing model. After multiple iterations, all weak learners are combined to form a strong final predictor $F_M(x)$, which provides the final load forecast.

*3.1.4 Long Short-Term Memory (LSTM)*

Long Short-Term Memory (LSTM) networks are specialized form RNNs designed to capture long-range dependencies in sequential data. LSTM models are effective because they maintain memory of past observations over time. This in return, makes them suitable for modeling temporal patterns such as daily, monthly, and yearly load cycles. Unlike traditional feedforward networks, LSTM units use gating mechanisms such as input, forget, and output gates to regulate the flow of information. This trains the network to learn when to retain or discard past information. Having this feature makes LSTMs efficacious at handling nonlinear, non-stationary, and seasonal load behaviors in load forecasting [34, 35]. The gate operations can be expressed in a matrix form as follows in Equation 3.12:

$$\begin{bmatrix} f_t \\ i_t \\ o_t \\ \tilde{c}_t \end{bmatrix} = \begin{bmatrix} \sigma \\ \sigma \\ \sigma \\ \tanh \end{bmatrix} \left( \begin{bmatrix} W_f & U_f \\ W_i & U_i \\ W_o & U_o \\ W_c & U_c \end{bmatrix} \cdot \begin{bmatrix} x_t \\ h_{t-1} \end{bmatrix} + \begin{bmatrix} b_f \\ b_i \\ b_o \\ b_c \end{bmatrix} \right) \quad (3.12)$$

The input at time step *t* can be $x_t \in R^d$, where d represents the number of input features (i.e., historical load values, temperature, time-of-day indicators). The LSTM cell maintains two main internal components [39]. The first being cell state $c_t$ which serves as the long-term memory, and the hidden state $h_t$ which is used for output and short-term memory. In Equation 3.12, $\sigma$ is the sigmoid activation function, and tanh is the hyperbolic tangent function. The matrices *W* and *U* are learnable weights associated with the current input and previous hidden state, respectively. Finally *b* denotes the bias terms. With the gates computed, the cell and hidden states are updated as follows:

$$c_t = f_t \odot c_{t-1} + i_t \odot \tilde{c}_t \qquad (3.13)$$

$$h_t = o_t \odot \tanh(c_t) \qquad (3.14)$$

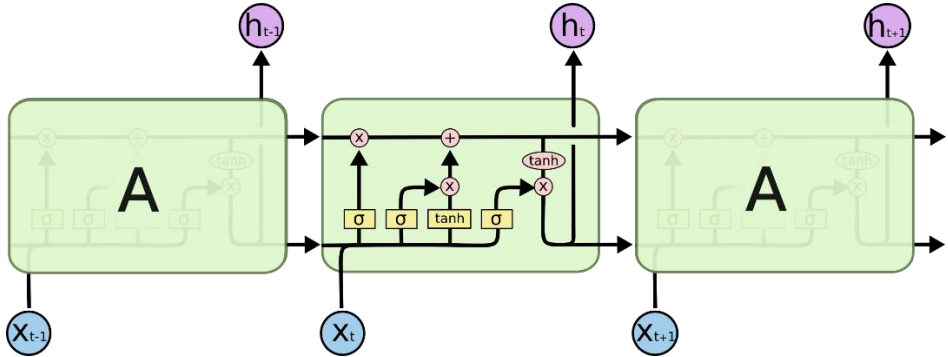

Figure 3.5: LSTM Network Architecture Across Time Steps [40].

Given three sequential time steps, *t-1, t,* and *t+1* as shown in Fig. 3.5. At each time step, the LSTM cell receives the input vector $x_t$ (which may contain features such as historical load values, temperature, and time of day) and the previous hidden state $h_{t-1}$. Within the cell, several gates control the flow of information. The forget gate determines how much of the previous memory $c_{t-1}$ should be retained, the input gate, and candidate memory update decide what new information should be added, and the output gate controls how much of the updated memory $c_t$ is exposed as the hidden state $h_t$. Finally, this hidden state is then conceded forward to the next time step and can be used to generate the predicted load. Thus, this is how the LSTM is able to selectively retain or discard information over time, making it well-suited for capturing both short-term fluctuations and long-term dependencies in electricity demand [35, 36, 40]. LSTM-based approaches have demonstrated improved forecasting accuracy in applications involving variable and seasonally influenced load profiles, particularly when applied within smart grid environments that require data-driven methods [37, 38]. Finally, LSTM models, through their sequence learning capability, sit well with the characteristics of load data, where demand is influenced by both immediate and historical conditions.

### *3.1.5 Hybrid ARIMA-SVM Approach*

The hybrid ARIMA–SVM approach adds the linear modeling capability of the Autoregressive Integrated Moving Average (ARIMA) model with the nonlinear regression capability of the Support Vector Machine (SVM). In this case, a hybrid ARIMA–Support Vector Regression (SVR) model is investigated. This modeling framework is based on the decomposition of a univariate time series into additive components, where the linear trend is modeled using ARIMA and the residual component is modeled using SVR. The basis behind this hybrid structure is that ARIMA is effective for capturing autocorrelations and linear dependencies in stationary time series data, but it does not perform well when the data exhibits nonlinearity, demand shifts, or load mean variations. On the other hand, SVR is a kernel-based regression method that is capable of modeling nonlinear relationships by mapping inputs into high-dimensional feature spaces using kernel functions as discussed earlier [48, 49, 50]. The final forecast in this hybrid approach is obtained by summing the predictions from both components. That is, the ARIMA model's estimate of the linear structure and the SVR model's estimate of the residual, nonlinear behavior. Moreover, training the SVR model on residuals that have been detrended by ARIMA reduces the complexity of the learning task, potentially lowering the risk of overfitting. It is also worth noting that hybrid approaches are not limited to ARIMA-SVM combinations; similar frameworks have been developed using other machine learning methods, such as artificial neural networks (ARIMA–ANN), to model residual patterns [51].

*3.2 Challenges and Limitations*

While the discussed ML approaches such as SVM, ANN, GBDT, and LSTM have improved the accuracy of load forecasting as compared to the classical approaches, challenges remain when applying these methods in practice. SVMs often struggle with scale as the dataset size increases. [29, 34, 36]. ANNs are sensitive to quality/quantity of the training data and require application-specific hyperparameter tuning. GBDTs are prone to overfitting and can become slow and too complex with large feature sets. Lastly, LSTM networks, though desired in load forecasting, may become unstable with load drifts and experience convergence issues. There is also the challenge of training ML models for newly built power grids due to limited data availability [39, 40, 41]. Despite the challenges, ML frames still offer great adaptability and thus, deserving of continued research and practical application.

CHAPTER 4

METHODOLOGY

*4.1 Analysis of Exogenous Factors Impacting Power Load*

Power load is influenced by several exogenous factors that are independent of the power system but directly affect the power demand [42]. In Arizona, the arid climate conditions amplify the influence of these factors, particularly during the summer months when cooling demand increases. Meteorological factors such as maximum and minimum temperatures, dew point, relative humidity, wind speed, atmospheric pressure, cloud cover, visibility, and ultraviolet (UV) index are can be accounted as relevant exogenous factors impacting power load [43]. Therefore, to quantify the relationship between each exogenous variable and the power load, the Pearson Correlation Coefficient is applied. This statistical measure evaluates the degree of linear association between two variables and is defined as:

$$r_{xy} = \frac{\sum_{i=1}^{n}(x_i - \bar{x})(y_i - \bar{y})}{\sqrt{\sum_{i=1}^{n}(x_i - \bar{x})^2} \sqrt{\sum_{i=1}^{n}(y_i - \bar{y})^2}} \qquad (4.1)$$

Where $x_i$ denotes the value of an exogenous environmental factor at time *i*, and $y_i$ denotes the denotes the corresponding power load at the same time. The means of *x* and *y* are represented by $\bar{x}$ and $\bar{y}$, respectively. The resulting coefficient $r_{xy}$ ranges between -1 and 1, indicating the degree to which the two variables move in tandem. Moreover, a value of $r_{xy}$ close to +1 suggests a strong positive linear relationship, indicating that as the exogenous variable increases, the power load tends to increase as well. Conversely, a value close to -1 implies a strong negative relationship, where an increase in the environmental factor has no correlation with the power load. A coefficient near 0 suggests little to no linear correlation between the variables. This correlation analysis forms an opening step in feature selection. Thus, the exogenous factors that have no contribution to the power load demand can be excluded and not used as a feature.

*4.2 Characteristics and Processing of Power Load Data*

Power load data represents the total amount of electrical energy consumed over a specific period, typically measured in megawatt-hours (MWh) or megawatts (MW). For this research, daily aggregated load data was obtained from the Arizona Public Service (APS) balancing authority via the U.S. Energy Information Administration (EIA). The load data entails the combined residential, commercial, and industrial energy demand within the APS territory and exhibits typical characteristics such as daily and seasonal variation, with noticeable peaks during summer months. To prepare the data for analysis, the original timestamp fields, which included the hour ending in local time format, were first converted into a steady date format. This was needed to enable merging with the corresponding daily weather records obtained from National Oceanic and Atmospheric Administration (NOAA) for the state of Arizona. The raw data on both ends needed further studying due to atypical entries such as empty or zero values not consistent with expected values.

*4.2.1 Handling of Atypical Data*

Upon studying the raw datasets, atypical values and some discrepancies were present. Neglecting these anomalies can skew the statistical characteristics of the data and thus, negatively impact the performance of forecasting models. To address such issues, this study applies two correction techniques: flat smoothing, which adjusts values based on local continuity in time, and periodic consistency adjustment, which leverages recurring daily patterns to identify and correct deviations.

Electric load typically varies in a gradual and continuous manner throughout the day. If there is a sudden and large deviation in the load value at a given time compared to adjacent time intervals, it may indicate an abnormal point. This is where flat smoothing is used. This method checks whether the difference between the current load and the previous or next time step exceeds an acceptable threshold:

$$|L(d,t) - L(d,t-1)| > \alpha(t) \qquad (4.2)$$

$$|L(d,t) - L(d,t+1)| > \beta(t) \qquad (4.3)$$

If such a deviation is detected, the value at that time is replaced with the average of its immediate neighbors:

$$L(d,t) = \frac{L(d,t-1) + L(d,t+1)}{2} \qquad (4.4)$$

Here, $L(d,t)$ denotes the load on day $d$ at time $t$, while $\alpha(t)$ and $\beta(t)$ are predefined thresholds based on acceptable intra-day variation. Power load also exhibits strong periodicity, especially daily patterns. Thus, values at the same hour across different days tend to follow a consistent trend. The periodic consistency adjustment detects abnormal points by comparing the load at a given time with the average load at that same time over recent days:

$$|L(d,t) - \mu(t)| > \delta(t) \qquad (4.5)$$

If the difference exceeds the threshold $\delta(t)$, the value is corrected using a boundary adjustment:

$$L(d,t) = \begin{cases} \mu(t) + \delta(t), & if\ L(d,t) > \mu(t) \\ \mu(t) - \delta(t), & if\ L(d,t) < \mu(t) \end{cases} \quad (4.6)$$

In this formulation, $\mu(t)$ is the historical mean load at time $t$, and $\delta(t)$ is the allowable deviation range.

### 4.2.2 Load Data Gap Correction

Given the strong daily periodicity in load behavior, a missing value at a specific time can be reasonably approximated using the load at the same time on the previous day or the following day. If both are available, their average is taken as the replacement value. When a value is missing at a specific time $t$ on day $d$, it can be estimated using values from the same time on adjacent days. If $L(d,t)$ is missing values from $d-1$ and day $d+1$ are available, the missing value is filled as:

$$L(d,t) = \frac{L(d-1,t) + L(d+1,t)}{2} \quad (4.7)$$

To illustrate this correction method, Fig. 4.1 simulates a scenario where load value at a specific time on day $d$ is estimated using the corresponding values from day $d-1$ and day $d+1$.

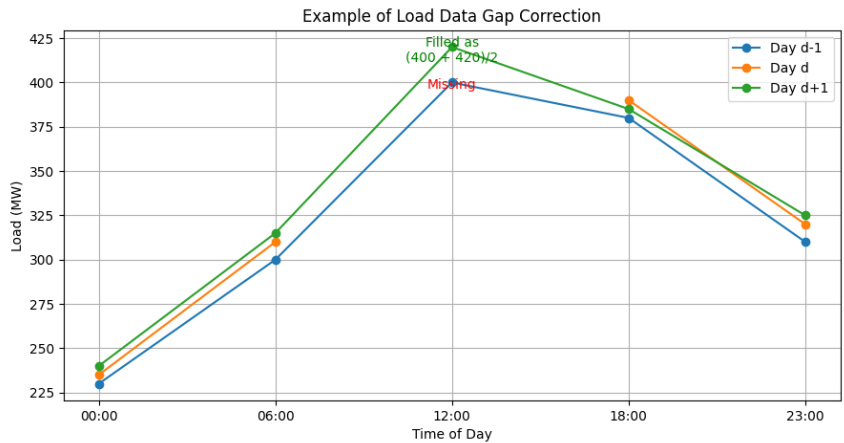

Figure 4.1: Load Data Gap Correction Using Periodic Consistency Adjustment.

In cases where only one adjacent day's data is available, the missing value can be directly substituted using the corresponding value from that day. Fig. 4.2 demonstrates this approach, where the load at a specific time on day *d* is filled using the value from day $d-1$.

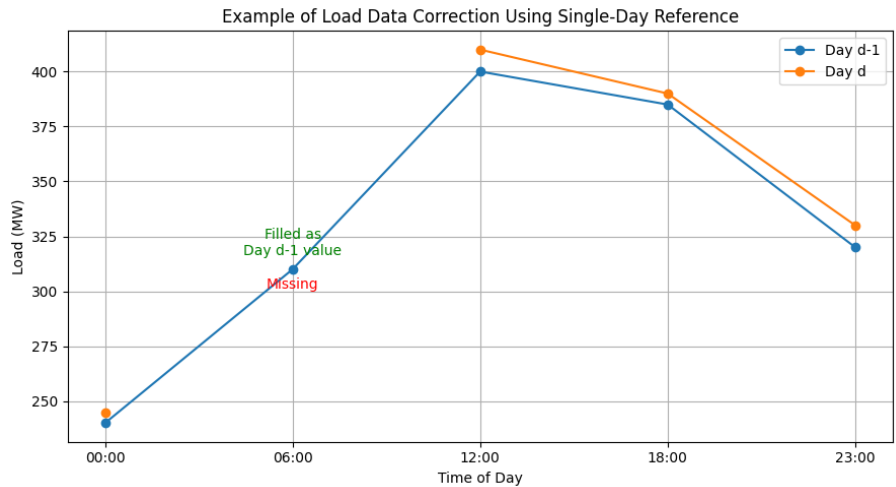

Figure 4.2: Load Gap Correction Using Single-Side Periodic Consistency Adjustment. This second simulation shows a case where the missing value (at 6:00 on day *d*) is filled using only the corresponding value from day $d-1$ as day $d+1$ is not available. In other words, this correction can act as a fallback strategy when only one adjacent day can be used.

*4.2.3 Data Normalization*

Following the atypical data handling and load data gap correction, the dataset is prepared for modeling by normalizing numerical variables and quantifying categorial inputs. This helps with preventing scale-related bias. The primary numerical variable is the power load, denoted as $\lambda$. To normalize the load data, a min-max transformation is applied, which scales all values to the interval [0, 1] based on the observed minimum and maximum. The transformation is defined as:

$$\lambda' = \frac{\lambda - \lambda_{min}}{\Delta \lambda} \quad (4.8)$$

Where $\Delta \lambda = \lambda_{max} - \lambda_{min}$. Here, $\lambda'$ is the normalized load value, and $\lambda_{max}$, $\lambda_{min}$ are the historical minimum and maximum load values. This normalization conserves the relative variation in load across time, while removing the influence of absolute magnitude. Moreover, to construe model predictions in actual load units, the reverse transformation is used:

$$\lambda = \lambda' \cdot \Delta \lambda + \lambda_{min} \quad (4.9)$$

In addition to numerical normalization, categorical features are also quantified. More specifically, holidays are encoded using a binary indicator: a value of 1 is assigned if the day is a holiday, and 0 otherwise. Finally, a simulation as shown in Fig. 4.3 demonstrates the process.

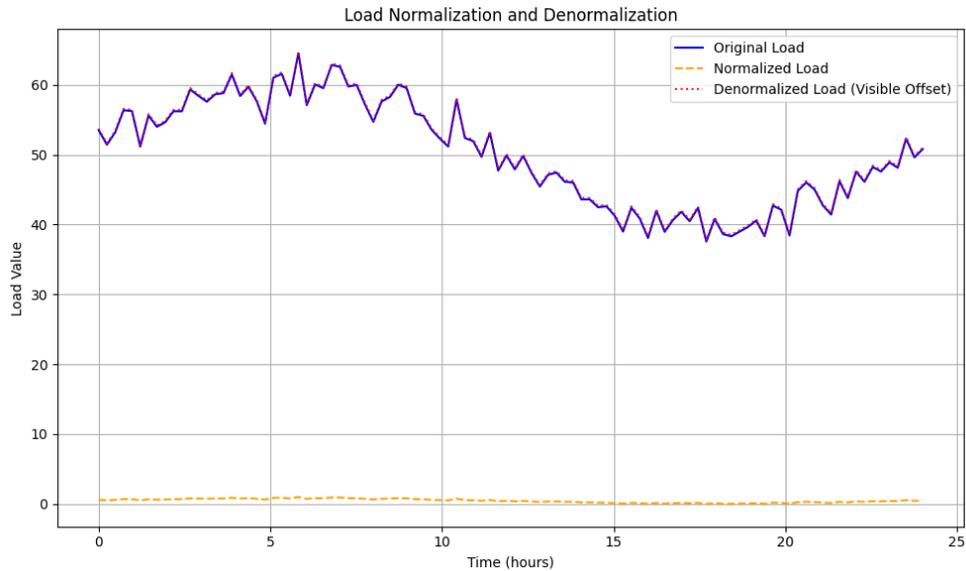

Figure 4.3: Visualization of Load Normalization and Denormalization with Offset.

As shown from the simulation in Fig. 4.3, the original, normalized, and denormalized load data over a 24-hour period is shown. After applying min-max transformation, the values are scaled between 0 and 1, which helps standardize the input for machine learning models. The denormalized load follows the original curve, thus, showing that the transformation preserves the data's structure. Additionally, to make the denormalized line visually distinguishable in Fig. 4.3, a slight offset was added. This minor adjustment does not reflect actual model output. It is used solely for clearer visualization. Without the offset, the denormalized curve overlaps the original load due to their near-identical values. Moreover, the offset applied in Fig. 4.3 is not part of the actual normalization or denormalization process and as such, has no influence on the quantitative results. Its sole purpose is to prevent visual overlap between the original and denormalized curves, which would otherwise appear as a single line due to similarity.

Table 4.1: Sample Load Data Showing Normalization and Denormalization Over a 24-Hour Period.

| Time (hours) | Original Load | Normalized Load | Denormalized Load |
|---|---|---|---|
| 0.0 | 53.53 | 0.5921 | 53.53 |
| 4.85 | 54.44 | 0.626 | 54.44 |
| 9.7 | 53.57 | 0.5938 | 53.57 |
| 14.55 | 42.47 | 0.1821 | 42.47 |
| 19.39 | 38.33 | 0.0284 | 38.33 |

Table 4.1 presents selected data points simulating the effect of min-max transformation and subsequent denormalization on load values over a 24-hour period.

The original load values represent simulated power demand with natural variation. These values are normalized to a 0–1 range to standardize the data for modeling purposes. The denormalized values, obtained by reversing the normalization process, closely match the original values, confirming the accuracy and reversibility of the transformation. Thus, this approach aimed to illustrate that the denormalization step accurately reconstructs the original data from the normalized values.

### 4.3 Forecast Evaluation Metrics

Several statistical metrics are used to quantify the accuracy of the predictions. Among the adopted are Mean Absolute Error (MAE), Mean Absolute Percentage Error (MAPE), Coefficient of Determination ($R^2$) and Root Mean Square Error (RMSE) [2]. Moreover, these evaluation metrics help assess how close the forecasted values are to the actual observed values. In this work, MAE, MAPE, $R^2$, and RSME are chosen as the primary forecast evaluation metrics. MAE is a linear measure, meaning, each error contributes equally to the final score, making it fit when equal weighting of all errors is desired. $R^2$, also known as the coefficient of determination, measures how well the predicted load values approximate the actual load values. MAPE, on the other hand, expresses the error as a percentage of the actual value, a normalized perspective that is used due to its simplicity. The mathematical expressions for MAPE and MAE, $R^2$, and RSME are presented in Equations 4.10, 4.11, and 4.12, 4.13 respectively.

$$MAPE = \frac{100}{n} \sum_{t=1}^{n} \left(\frac{A_t - F_t}{A_t}\right) \qquad (4.10)$$

$$MAE = \frac{1}{n} \sum_{t=1}^{n} |A_t - F_t| \qquad (4.11)$$

$$R^2 = 1 - \frac{\sum_{t=1}^{n}(A_t - F_t)^2}{\sum_{t=1}^{n}(A_t - \bar{A})^2} \tag{4.12}$$

$$RSME = \sqrt{\frac{1}{n}\sum_{t=1}^{n}(A_t - F_t)^2} \tag{4.13}$$

In all evaluation metrics, $A_t$ represents the actual value of the electrical load at time $t$, $F_t$ is the forecasted (predicted) value of the load at time $t$, $\bar{A}$ is the mean of all actual load values over the evaluated time period, and $n$ is the total number of data points / time step observations used in the evaluation metric.

## *4.4 Feature Selection*

Before working on any model implementation, it is worth studying the features that have large influence on power load consumption. The electricity demand data obtained from the EIA required initial preprocessing on the load data. The EIA provides open-access, region-specific electricity demand data through its grid monitor platform. The dataset, obtained from the Arizona Public Service (APS) balancing authority, included daily demand values along with time stamps labeled by hour ending and local time notation. The timestamps were first converted into a standardized date format to enable merging with the NOAA weather data. Data normalization techniques were performed on the load dataset as discussed in section 4.2.3. Atypical entries were dealt with in accordance with the techniques discussed in 4.2.1. Given that the EIA provides demand in megawatt-hours (MWh), no unit conversion was necessary. However, some days included outliers or anomalies. These were identified through inspection of the dataset. Moreover, the NOAA weather data was studied. Although the dataset included many weather-related variables, not all variables moved in tandem with the load data. Additionally, using/keeping all features indiscriminately increases computational burden and training time. As stated in 4.1, not all exogenous factors impact the power load meaningfully. Thus, keeping non-correlative data can complicate the models and even worse, degrade the forecasting performance. To solve this problem, Pearson coefficient method, as discussed in 4.1 was adopted. It was important to mention that before analysis, normalization is performed first, then Pearson similarity is conducted.

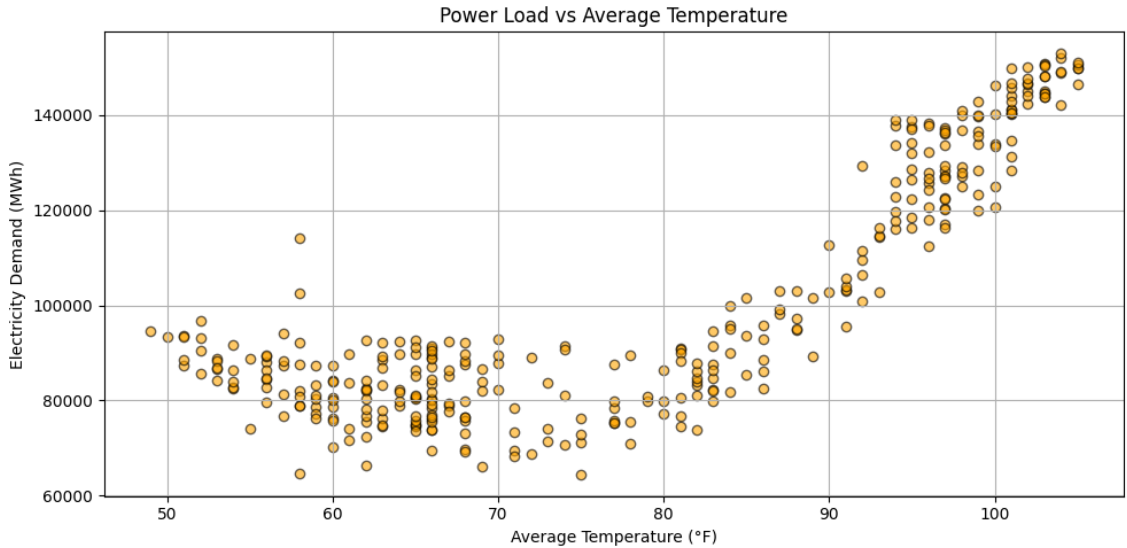

Figure 4.4: Scatter Plot of the Relationship Between Average Daily Ambient Temperature and Corresponding Electricity Demand Levels Over Time.

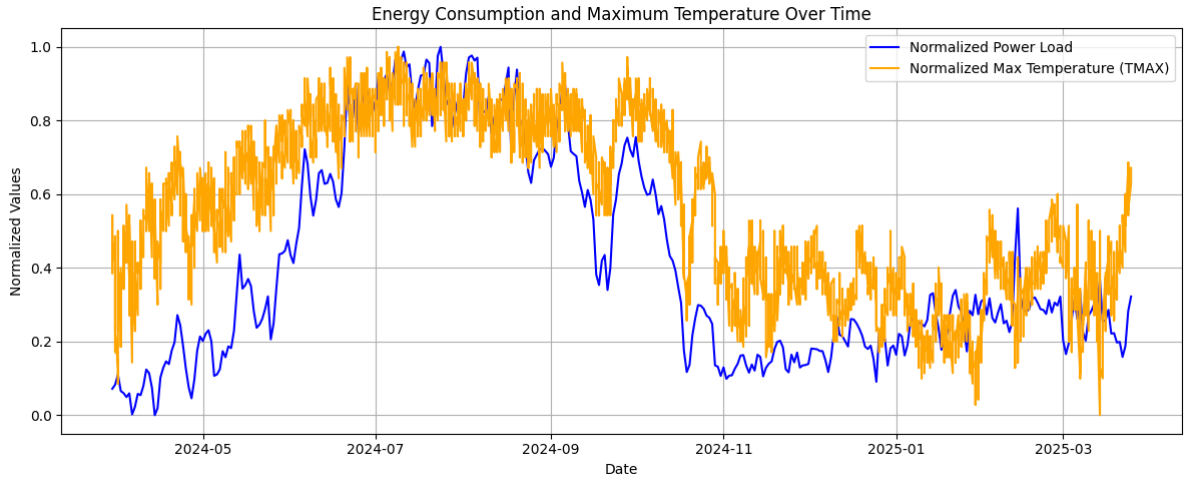

Figure 4.5: Observed Relationship Between Average Daily Temperature and Normalized Electricity Demand Based on Historical Weather and Load Data.

Fig. 4.4 presents a scatter plot that denotes the relationship between average daily temperature and electricity demand. Each point in the plot represents a single day, with the *x-axis* showing the average ambient temperature and the *y-axis* representing the corresponding electricity demand in megawatt-hours (MWh). A visible upward trend suggests a positive correlation: meaning, as the temperature increases, electricity demand tends to rise as well. Moreover, Fig 4.5 presents a plot comparing normalized daily electricity demand with maximum temperature (TMAX) over the same period. Both variables have been scaled to a common range to allow for direct visual comparison. The plot shows a strong seasonal pattern: peaks in TMAX make parallel with increases in electricity demand, especially during the hotter months. Given this, the influence of rising temperatures on energy consumption is observed.

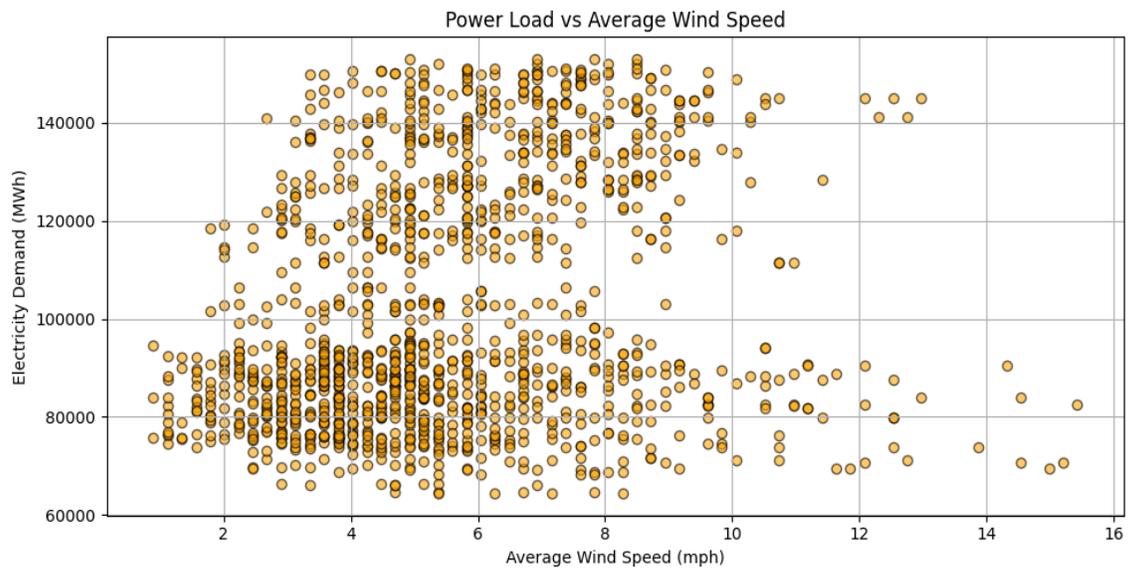

Figure 4.6: Scatter Plot Showing Lack of Relationship Between Average Wind Speed and Electricity Demand.

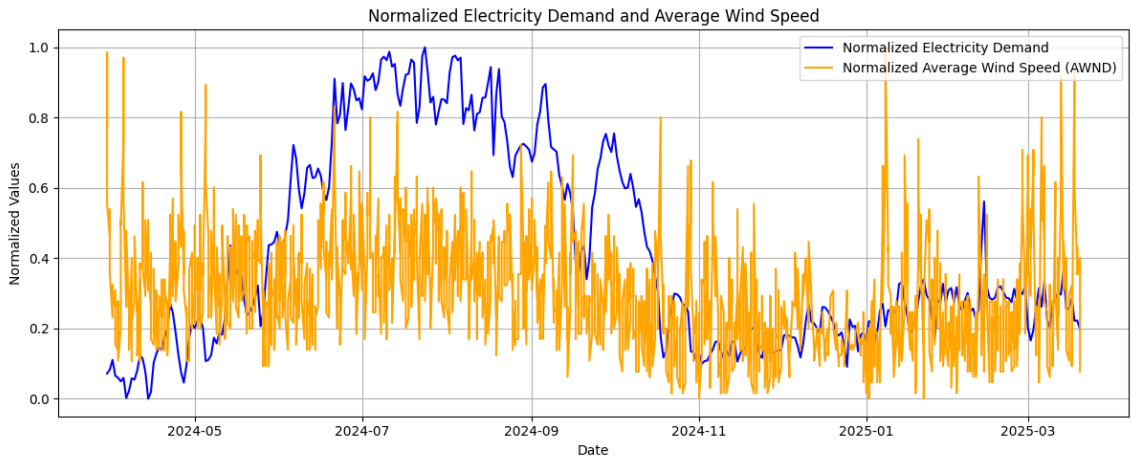

Figure 4.7: Normalized Electricity Demand Compared with Average Wind Speed.

Fig. 4.6 presents a scatter plot denoting the relationship between average wind speed and electricity demand. Each point represents a single day's observation, where the horizontal axis corresponds to average wind speed and the vertical axis shows the electricity demand for that day. The data points appear scattered with no clear pattern or directional trend. This ultimately indicates that wind speed does not have a meaningful influence on electricity consumption in Arizona during the observed period. The lack of correlation proposes that wind speed may not be a valuable feature for forecasting and thus, can be excluded from model training to reduce complexity without sacrificing accuracy.

To complete the feature selection analysis, the Pearson correlation coefficient will be used to quantify the relationships between electricity demand and other exogenous variables available in the dataset. Finally, the results of the Pearson correlation coefficient will determine the input features which will be implemented to the ML models.

CHAPTER 5

IMPLEMENTATIONS

*5.1 Baseline ARIMA Model*

As discussed in chapter two, before applying the ARIMA model, the time series data must be stationary, meaning, its statistical properties do not change over time. Thus, stationarity is assessed using the Dickey-Fuller test; if the p-value is below 0.05, the series is considered stationary. If not, differencing is applied to remove trends or seasonality. In this case, the load demand time series became stationary after first-order differencing, indicating that the differencing parameter $d$ is equal to 1. Moreover, the autoregressive parameter $p$ is determined using the partial autocorrelation function (PACF) plot, while the moving average parameter $q$ is selected based on the autocorrelation function (ACF) plot.

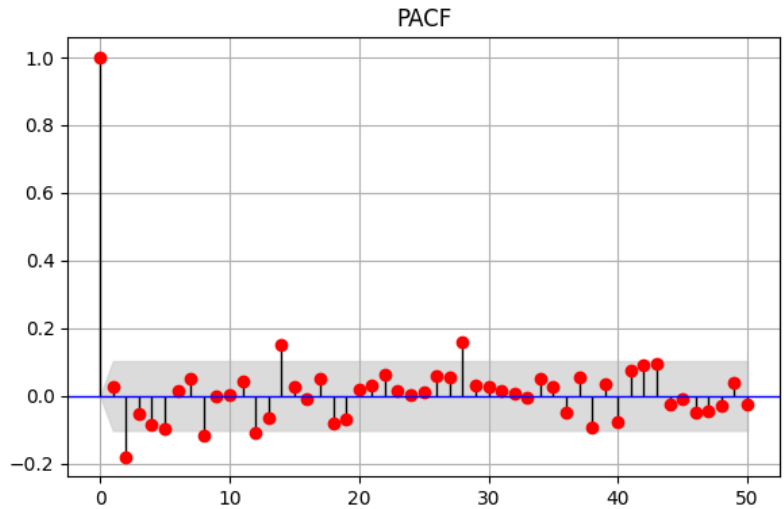

Figure 5.1: PACF plot showing lag significance for autoregressive order selection after differencing.

Shown in Fig. 5.1, the PACF plot denotes a sharp decline after lag 1, which indicates that most of the autoregressive structure is seized by the first lag. This shows that adding more AR terms beyond lag 1 would contribute little to improving the load model.

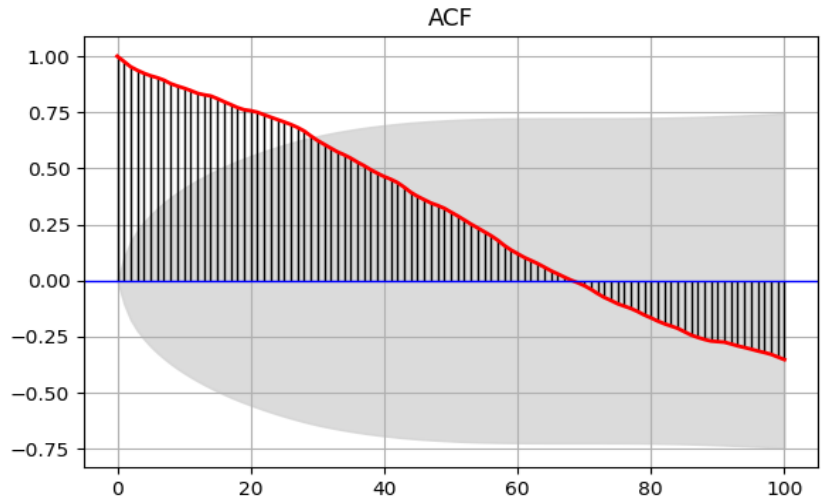

Figure 5.2: ACF plot showing persistence decay pattern in original load series for moving average order selection.

As PACF plot showcases the correlation between the time series and its lag, it helps in determining which lag is needed for the AR term. Moreover, $q$ is determined from the ACF plot as shown in Fig. 5.2. The plot depicts a gradual decay and therefore, the describes the number of MA terms needed to remove remote auto-correlation in the stationary time series. With these observations and after a first-order differencing, the ARIMA model was trained on the stationary load data. The results of this baseline ARIMA model are explained in chapter six, section 6.2. It is important to mention that this baseline ARIMA model only relies on past load values and does not consider exogenous factors such as weather data, holidays, or other calendar events. It operates purely on the time series structure without incorporating any exogenous variables. Despite this limitation, the model provides a good starting point and as such, serves as a reference for evaluating the performance of more complex models, including those built using ML-based techniques.

*5.2 Hybrid ARIMA-SVM Model*

The hybrid ARIMA-SVM model is implemented in two stages, as it combines the strengths of linear time series forecasting with nonlinear machine learning regression. In the first stage, an ARIMA model is used to capture the linear temporal dependencies in the load data. Once the ARIMA model is fit to the training data, as explained in section 5.1, it produces forecasted values based on historical load values. These values are then subtracted from the actual observed values to compute the residuals, which contain the nonlinear components of the series that were not explained by ARIMA. In the second stage, the residual series is then modeled using a Support Vector Regression (SVR) model. Moreover, SVR is a variant of the Support Vector Machine (SVM) algorithm, used for regression, as discussed in chapter three. This attempt is to find a function that estimates the relationship between the input features and the residuals within a specified margin of tolerance, known as the epsilon-tube, as explained in chapter three. However, unlike linear regression, SVR uses kernel functions to map the input data into a higher dimensional space to capture nonlinear load relationships. In this implementation, a radial basis function (RBF) kernel was chosen. This is because RBF methodology can generalize well on large datasets such as the load dataset used in this research. For the SVR model, the input features consist of lagged load values and lagged residuals. A summary of the engineered features is provided in Table 5.1 below:

Table 5.1: Summary of Engineered Time-Series Features for Forecasting for hybrid ARIMA-SVM Model

| Feature Name | Description |
| --- | --- |
| load_lag1 | The previous day's load as input to SVR for learning temporal dependencies. |
| rolling_3 | A 3-day rolling average of load, capturing short-term trends. |
| rolling_7 | A 7-day rolling average of load, representing weekly consumption behavior. |
| day_of_week | Integer encoding of the weekday (0 = Monday, 6 = Sunday), capturing cyclic weekly behavior. |

As shown in table 5.1 above, the features were selected based on their ability to reflect short-term patterns and weekly seasonality in electricity demand. This was considered without introducing external dependencies. Additionally, the lag and rolling averages for the SVR learn the variations in the residual structure and the weekday encoding helps with recurring load behavior across the week. Taken together, they form the complete input set for the nonlinear regression stage. As discussed earlier, with the ARIMA model capturing the linear components and the SVR trained to model the residual nonlinearities in the load data. The forecasting results produced by the hybrid ARIMA-SVM model, along with its performance evaluation, will be presented in the following chapter.

*5.3 Gradient Boosting Decision Trees*

As discussed previously, GBDTs are scalable machine learning models that work well for structured data. This is what makes them a good candidate in load forecasting scenarios. For this study, they were applied to daily load demands in Arizona using the historical load demand collected from years 2020 to 2025. The load demand data—specifically for APS (Arizona Public Service), the state's largest electric utility, was obtained from the U.S. Energy Information Administration (EIA) via an application programming interface (API) call. The historical load data was then preprocessed and normalized, as discussed in chapter four, before being used for model training. Since electricity demand is closely linked to both weather and temporal factors, the model incorporated temperature as the primary feature, along with engineered variables such as previous-day load (lag), rolling averages, and day-of-week indicators. The selection of temperature as the main weather-related input was based on a Pearson correlation analysis conducted between load and various meteorological exogenous variables. The analysis confirmed a strong positive correlation between rising temperatures and increased electricity demand, especially during warmer months. The results of this correlation study are presented in the following chapter. Finally, the actual GBDT implementations were carried out using XGBoost and LightGBM in the following sections.

*5.3.1 XGBoost Algorithm*

To implement the XGBoost algorithm for short-term load forecasting, the first step involved preparing the dataset, which included daily load demand via Arizona Public Service (APS) balancing authority from January 2020 through January 2025. The load data was sourced from the EIA via an API call and the resulting .csv file was saved for preprocessing. Next, the corresponding related weather data obtained from NOAA also underwent preprocessing, as explained in the earlier chapter. Moreover, the weather data contained features such as average temperature (TAVG), maximum temperature (TMAX), minimum temperature (TMIN), and other related meteorological variables. Pearson correlation coefficient (PCC) was formed on the datasets so determine which features had the highest correlations with the load demand. The results of PCC are explained in the following chapter. Moreover, from the dataset, the temperature was computed using a preference for average temperature (TAVG), or if unavailable, the average of the maximum and minimum temperatures. Once merged with the load dataset, any missing values were dropped to maintain data consistency. Finally, to improve the model's ability to recognize, but more importantly, learn from the load patterns and temporal trends, several time-series features were engineered from the raw load data. These features aid the model account for short-term variations, weekly cycles, and the influence of recent demand on future values. A summary of the engineered features is provided in Table 5.2 below:

Table 5.2: Summary of Engineered Time-Series Features for Forecasting for XGBoost.

| Feature Name | Description |
| --- | --- |
| load_lag1 | The previous day's load, to account for day-to-day demand continuity. |
| rolling_3 | A 3-day rolling average of load, capturing short-term trends. |
| rolling_7 | A 7-day rolling average of load, representing weekly consumption behavior. |
| day_of_week | Integer representation of the weekday, capturing recurring weekly patterns. |

After feature engineering, input selection, and entries processing, the dataset was split into features (*x*) and target variable (*y*), which was the actual load in megawatt-hours. To simulate a realistic short-term forecasting scenario, the final 7 days of the dataset were held out as the test set, and the rest of the data was used as the training set. This method simulates how forecasts are made in real power applications, using past data to predict future demand. Moreover, the XGBoost model was then initialized with a manually tuned set of hyperparameters, which had been identified as effective through prior experimentation. This produced reasonably good results, but to further improve performance, a custom hyperparameter optimizer was introduced. The optimizer sampled random combinations of hyperparameters and validated each one using three-fold time series cross-validation. For every trial, the average MAPE across the three validation folds was calculated. The combination which resulted in the lowest average MAPE was selected as the chosen configuration. The hyperparameter configuration obtained through this random optimization technique is tabulated in Table 5.3 below:

Table 5.3: XGBoost Hyperparameters from Customized Random Search-Based Optimizer.

| Hyperparameter | Value | Description |
|---|---|---|
| n_estimators | 300 | Number of boosting rounds. |
| max_depth | 5 | Maximum depth of each tree. |
| learning_rate | 0.06 | Step size shrinkage to prevent overfitting. |
| subsample | 0.8 | Fraction of data used per tree. |
| colsample_bytree | 0.8 | Fraction of features used per trees. |
| random_state | 42 | Seed for reproducibility. |

The forecasting results produced by the XGBoost model, along with its performance evaluation, will be presented in the following chapter. Additionally, for comparative purposes, results from the XGBoost model without the customized hyperparameter optimizer will also be showcased to discuss the improvements obtained through the optimization process.

### *5.3.2 LightGBM Algorithm*

LightGBM is a fast, distributed, high-performance gradient boosting framework based on decision tree algorithms. It is designed to be efficient in both memory usage and computation. This is especially evident with large datasets and many features. Compared to traditional GBDT implementations, LightGBM is optimized with techniques such as histogram-based decision tree learning, leaf-wise tree growth, and parallel training. In this study, LightGBM was implemented to forecast daily electricity demand for the APS balancing authority. The model was trained using historical load demand data from years 2020 through 2025, collected via EIA API call. The load dataset was then aligned with daily weather data for Arizona obtained from NOAA. As discussed earlier, a PCC was created to identify exogenous feature correlations with the historical daily load data. The results of PCC are explained in the following chapter. The dataset was structured to enhance temporal awareness in the model by including both the target variable and several supporting features. The target variable was *load_mwh*, representing the actual daily electricity consumption. Several engineered features were introduced capture recent trends and cyclical patterns in the data. A summary of the engineered features is provided in Table 5.4 below:

Table 5.4: Summary of Engineered Time-Series Features for Forecasting using LightGBM.

| Feature Name | Description |
| --- | --- |
| load_lag1 | The previous day's load, for day-to-day demand continuity. |
| rolling_3 | 3-day rolling average of load demand. |
| rolling_7 | 7-day rolling average of load demand. |
| day_of_week | Daily index from 0 (Monday) to 6 (Sunday). |

As shown in Table 5.4, the *load_lag1* feature captures the previous day's load, thus the model learns the recent consumption behavior. The *rolling_3* feature introduces a 3-day rolling average which is used to represent short-term trends in demand. The *rolling_7* feature provides a 7-day rolling average, that captures the wider weekly consumption patterns. Finally, the *day_of_week* feature encodes the weekday as an integer from 0 (Monday) to 6 (Sunday), thus making the model learn and generalize typical weekly cycles.

Similar to the XGBoost approach, to improve the forecasting performance of the LightGBM model, a custom hyperparameter optimization routine was developed. Rather than relying solely on manual tuning, or default researched values, this approach involved programmatically testing multiple combinations of hyperparameters to identify those that produced the most accurate forecasts on the test set. The custom optimizer randomly sampled from predefined ranges of main hyperparameters such as *n_estimators*, *learning_rate*, *max_depth*, *subsample*, and *colsample_bytree*. For each trial, a model was trained using the sampled configuration, and its predictions were evaluated using the MAPE. The hyperparameter configuration obtained through this optimization technique is tabulated in Table 5.5 below:

Table 5.5: LightGBM Hyperparameters from Customized Random Search-Based Optimizer.

| Hyperparameter | Value | Description |
| --- | --- | --- |
| n_estimators | 300 | Number of boosting rounds. |
| max_depth | 5 | Maximum depth of each tree. |
| learning_rate | 0.06 | Step size shrinkage to prevent overfitting. |
| subsample | 0.4 | Fraction of data used per tree. |
| colsample_bytree | 1.0 | Fraction of features used per trees. |
| random_state | 50 | Seed for reproducibility. |

The forecasting results produced by the LightGBM model, along with its performance evaluation, will be presented in the following chapter. Additionally, for comparative purposes, results from the LightGBM model without the customized hyperparameter optimizer will also be showcased to discuss the improvements obtained through the optimization process.

*5.4 Recurrent Neural Networks*

Recurrent Neural Networks (RNNs) are a type of neural network architecture specifically designed to handle sequential data, such as time series, where the order and context of previous values are important in predicting future values. Unlike traditional feedforward networks, RNNs incorporate a feedback loop that allows information to persist across time steps. Thus, the model can learn temporal dependencies. This feature makes RNNs great for load forecasting, where current values are often influenced by patterns from previous days. While RNNs are conceptually powerful, they can struggle with learning long-term dependencies due to issues like vanishing gradients. To resolve this issue, more advanced architectures have been developed, including Long Short-Term Memory (LSTM) networks and Gated Recurrent Units (GRUs). Both LSTMs and GRUs introduce internal gating mechanisms that help the network retain relevant information over longer periods while discarding irrelevant inputs.

In this study, both LSTM and GRU networks were implemented to forecast daily load demand. These models were trained using historical load data, along with weather data, to better capture both temporal and exogenous influences. LSTM and GRU models were configured using a recursive forecasting approach, where each day's prediction was used as input for the next. The accuracy of the RNN models will be discussed and explored in the following chapter.

*5.4.1 Long Short-Term Memory Architecture*

To implement the LSTM model, a sequence-based approach was used using historical daily load data. The input was structured as 14-day rolling windows which allowed the model to learn demand patterns over a consistent timeframe. Each input sequence included features such as daily load, temperature, lagged load values, and rolling averages. The dataset was normalized to improve training stability, and samples were reshaped to reflect a three-dimensional format suitable for sequence modeling: samples, timesteps, and features. The LSTM architecture comprised of two sequential LSTM layers to capture both immediate and delayed temporal dependencies in the input sequences. The first LSTM layer contained 50 memory units and was configured to return the full sequence output, which was then passed to the second LSTM layer. This second layer processed the sequence further and helped improve the understanding of multi-day patterns. Additionally, dropout layers with rates of 0.3 and 0.2 were placed after each LSTM layer, respectively. This was done to reduce the risk of overfitting. These dropout layers randomly deactivate a portion of neurons during training, which contributes to better generalization when the model is applied to unseen data. Moreover, the model concluded with a dense output layer that produced a single load prediction for each input sequence. It was compiled using the Adam optimizer with a learning rate of 0.005 and used the mean absolute error as the loss function. To improve training efficiency and avoid overfitting, an early stopping mechanism was implemented. This feature monitored the training loss and halted the process if no further improvement was observed over several epochs, while restoring the performing model weights. Overall, this architecture was implemented to maintain depth and simplicity, thus aiding the model to extract and learn useful patterns from temporal sequences while maintaining generalization through regularization and early stopping. This architecture performed well for short-term forecasting tasks, where daily fluctuations and short-term memory effects significantly influence load behavior.

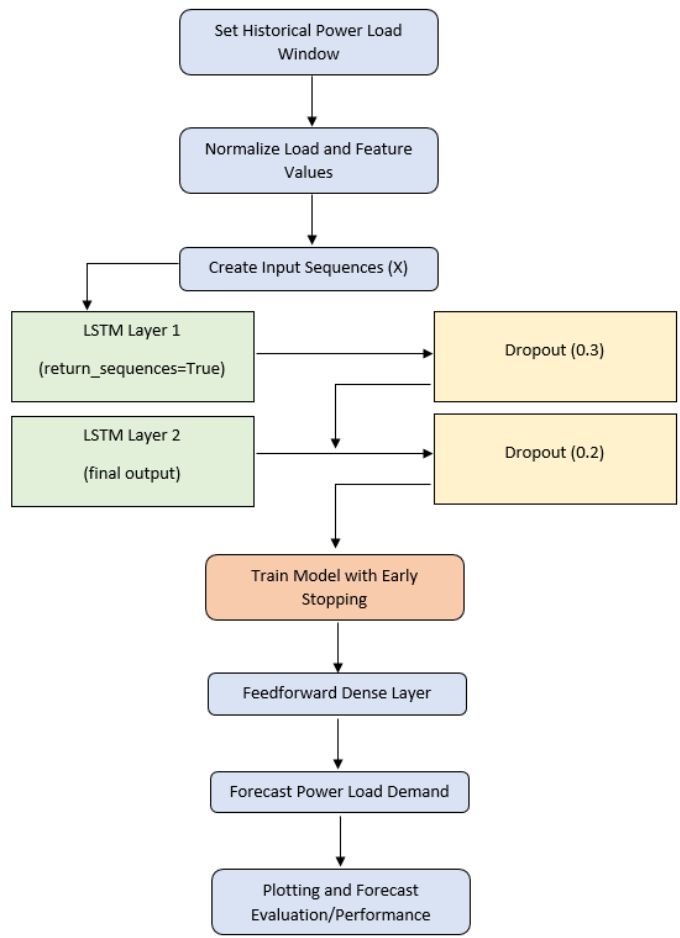

Figure 5.3: LSTM Load Forecasting Model Architecture with Early Stopping and Evaluation.

Figure 5.3 presents a visual representation of the implemented LSTM load forecasting model, corresponding directly to the Python code used in this study. The diagram outlines each stage of the modeling pipeline, beginning with the selection of a historical input window and the normalization of both power load and auxiliary features such as temperature and temporal indicators. This is followed by the creation of 14-day input sequences, which were structured into a three-dimensional array format compatible with LSTM input requirements. The model architecture depicted in the figure mirrors the code implementation, starting with two stacked LSTM layers. The first LSTM layer is configured with *return_sequences=True* to pass full sequential information to the second LSTM layer, both of which are followed by dropout layers with rates of 0.3 and 0.2, respectively. These layers, shown in Fig 5.3, aid in preventing overfitting by randomly deactivating units during training. The output from the LSTM layers is passed to a dense layer that produces the final load prediction for each input sequence. Additionally, the figure includes a training block that incorporates early stopping. The feature is implemented using the *EarlyStopping* callback. This monitors the training loss and halts the process if no improvement is detected over multiple epochs, restoring the better performing weights. In summary, Figure 5.3 depicts the full scope of the LSTM model design and training pipeline, thus providing a structural overview of the implementation. The performance outcomes resulting from this LSTM architecture will be discussed in the following chapter.

*5.4.2 Gated Recurrent Unit*

The Gated Recurrent Unit (GRU)-based load forecasting model was implemented as an alternative sequence modeling approach to the LSTM, with the goal of capturing short-term dependencies in the load behavior. The architecture followed a similar preprocessing pipeline, starting with historical daily load data which was improved through feature engineering. This included a 3-day lag for both load and temperature, along with categorical indicators such as the day of the week and weekend flags. These features were selected to account for recent temporal trends and behavioral patterns that influence load consumption. The input data was normalized using MinMax scaling to make consistent feature ranges. Moreover, a three-dimensional array with a single timestep was created for compatibility with GRU input expectations. The GRU model itself was constructed using two sequential GRU layers. The first GRU layer was configured with 100 units and set to return sequences, which allowed the second GRU layer, containing 50 units, to receive the full sequence information. These two layers were designed to extract temporal relationships in the input features, particularly focusing on recent variations in load and weather conditions. Dropout layers with rates of 0.3 and 0.2 were added after each GRU layer to reduce overfitting and improve the overall model generalization. A final dense layer was used to produce the predicted load value. The model was compiled using the Adam optimizer with a learning rate of 0.005 and trained using mean absolute error (MAE) as the loss function.

To prevent overfitting during training, an early stopping mechanism was created. This feature monitored the training loss and terminated the learning process if no improvement was observed over 10 consecutive epochs, while also restoring the better performing weights. The model's performance was evaluated on the final 7 days of the dataset, with predictions compared against actual load values using mean absolute percentage error (MAE) and mean absolute percentage error (MAPE). Overall, the GRU implementation provided a lighter-weight alternative to LSTM with a reduced number of internal parameters. The results and comparative performance of the GRU model will be presented and analyzed in the subsequent chapter.

# CHAPTER 6

# EXPERIMENTS & RESULTS

## 6.1 Pearson Correlation Results

In this study, Pearson Correlation Coefficient (PCC) was applied to quantify how various weather-related features relate to average energy consumption. PCC is a statistical measure to evaluate the strength and direction of the linear relationship between two continuous variables. It returns values between -1 and 1, where values closer to 1 or -1 indicate strong positive or negative correlations, respectively, while values near 0 suggest a weak or no linear relationship.

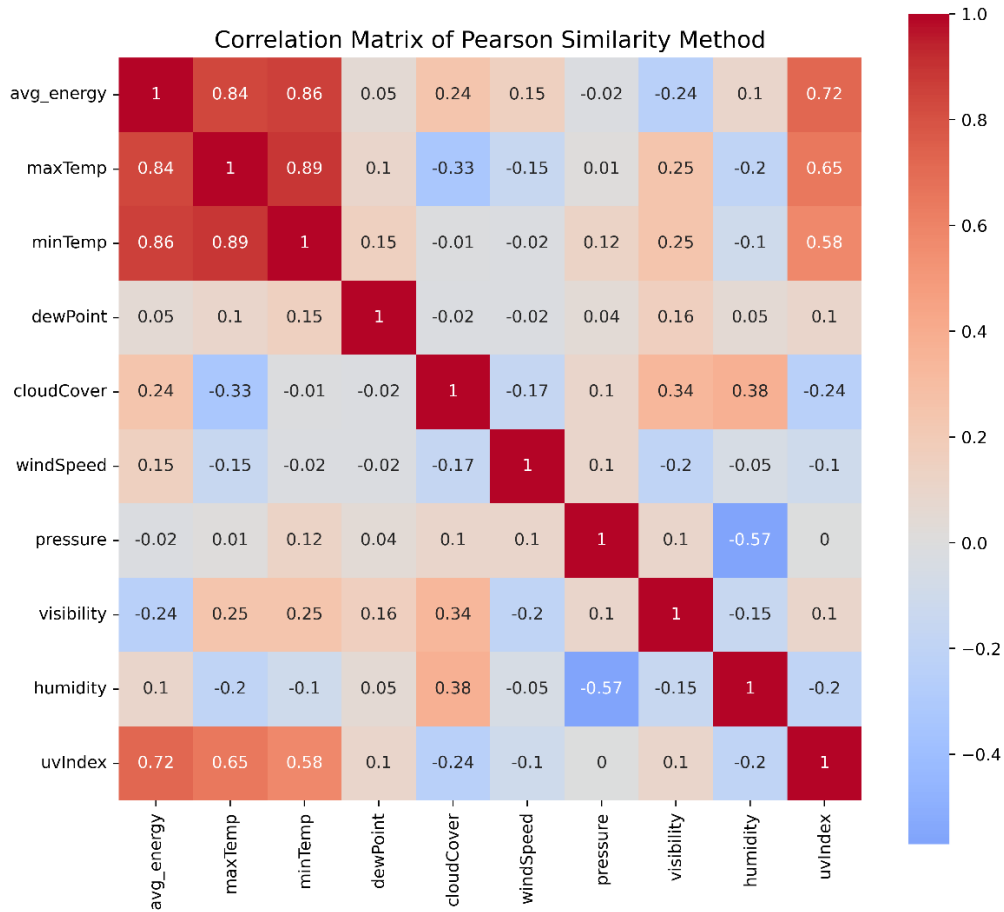

Figure 6.0: Pearson Correlation Matrix of Average Energy and Weather Variables.

*6.2 Pearson Correlation Timeframe and Data Results*

The results of the PCC analysis are shown in Fig 6.0. This matrix presents the correlation values for each pair of variables, including average energy demand and meteorological data obtained from NOAA for the state of Arizona. Moreover, it can be seen that the variable *avg_energy* shows the strongest positive correlations with *minTemp* (0.86), *maxTemp* (0.84), and *uvIndex* (0.72). These findings indicate that given the arid climate of Arizona, as the temperature gets warmer, the electricity load experiences an increased demand. Moreover, same goes true with the variable *uvIndex* as well. Sunny days typically signify a higher UV index, which in return, can also mean higher overall temperatures. This is consistent with climate-related demand shifts, as higher outdoor temperatures typically result in increased use of air conditioning systems. In contrast, other weather-related variables such as *dewPoint*, *cloudCover*, *windSpeed*, and *pressure* display weaker correlations with energy demand, suggesting a less direct influence on short-term load behavior.

Additionally, variables such as *humidity* and *visibility* show small negative associations with energy usage, with *visibility* having a correlation coefficient of -0.24. Although the magnitude of these values is modest, they do not cause a significant impact on the load demand behavior. The correlation results steered the selection of relevant exogenous variables for use in the machine learning forecasting models. Finally, this experimentation helped include inputs that add useful variation to the data without introducing excess noise or making the model overly complicated. This step was important for improving how well the models performed during actual prediction.

## 6.3 Baseline ARIMA Model Results

The baseline ARIMA model was applied using the historical daily load data, without additional exogenous inputs. This model relied on the internal structure of the load series, including trends and seasonality, to generate short-term forecasts. The performance of this univariate model was measured as follows:

Table 6.1: Baseline ARIMA Model Forecast Performance.

| Model | MAE | MAPE |
| --- | --- | --- |
| ARIMA | 3878.2 (MWh) | 4.03% |

Figure 6.1 illustrates the comparison between actual daily load demand and the forecasted values by the baseline ARIMA model over a 7-day period. The blue line represents the actual observed load, while the orange dashed line shows the predicted values generated by the ARIMA model.

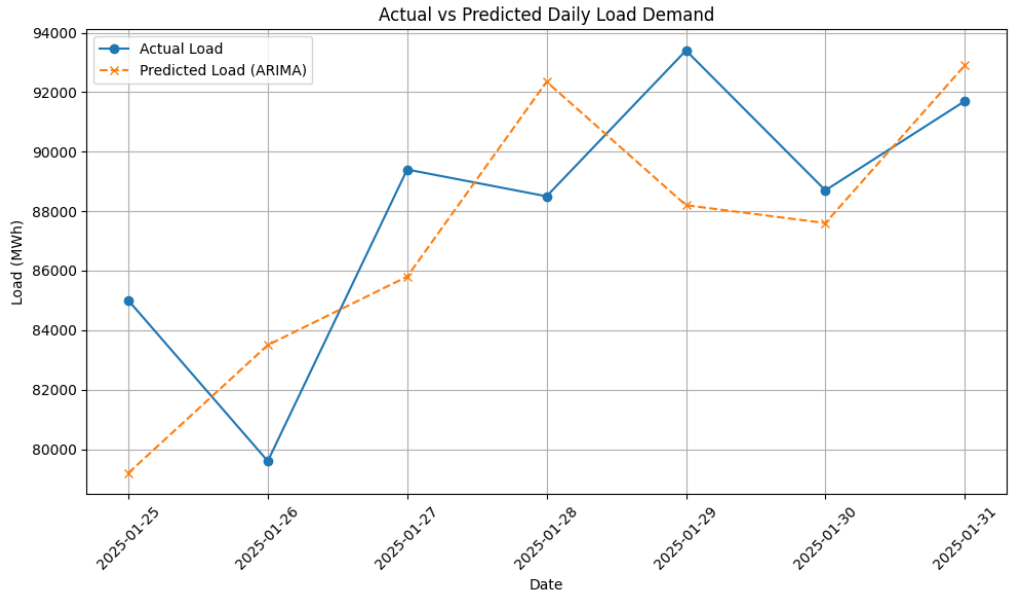

Figure 6.1: Baseline ARIMA Forecast vs Actual Daily Load Demand.

As shown in Fig. 6.1, The baseline ARIMA model, which was trained using only historical load data without any external features, follows the general trend of the actual load values. While the model underestimates some peak values and slightly overestimates on certain days, the overall forecast remains close to the observed demand. The prediction errors are modest, with a mean absolute percentage error (MAPE) of 4.03% and a mean absolute error (MAE) of 3878.2 MWh. These results indicate that the model captures the underlying patterns in the data with reasonable accuracy, serving as a solid reference point for evaluating more advanced forecasting models that incorporate exogenous variables.

*6.4 Hybrid ARIMA Model Results*

The results of the Hybrid ARIMA-SVM model is presented in Fig 6.2 and show the model's capacity to capture the general load demand trend across the 7-day forecast horizon for short-term load forecasting. In the plotted comparison between the actual and predicted load values, it can be observed that the predicted values follow the overall direction of the actual load curve reasonably well, especially on days where there are no abrupt fluctuations. Moreover, the hybrid ARIMA-SVM model outperformed the baseline ARIMA, which lacked the capacity to account for nonlinear dependencies. The final forecasting results of the hybrid ARIMA-SVM model are tabulated in Table 6.2 below:

Table 6.2: Hybrid ARIMA-SVM Model Forecast Performance.

| Model | MAE | MAPE |
| --- | --- | --- |
| Hybrid ARIMA-SVM | 1857.14 (MWh) | 2.09% |

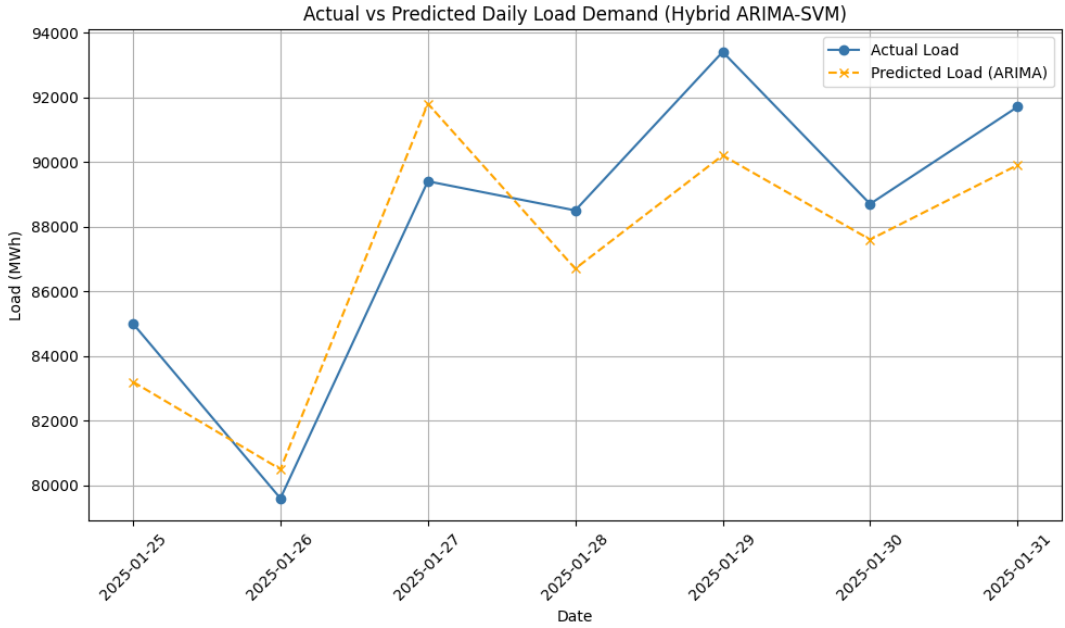

Figure 6.2: Hybrid ARIMA-SVM Forecast vs Actual Load Values

*6.5 XGBoost Model Results*

The results of the XGBoost algorithm are illustrated in Figures 6.3 and 6.4, which present the model's predicted daily load values against the actual observed load values over the selected 7-day period. Figure 6.2 shows the output of the XGBoost model using initial hyperparameter settings. In this case, the predictions fail to follow some of the sharper movements in the actual load, especially around days with more noticeable peaks and drops. This limited alignment is reflected in the error metrics, where the model produced a mean absolute error (MAE) of 4112.51 and a mean absolute percentage error (MAPE) of 4.67%, both of which were worse than those of the baseline ARIMA model.

Figure 6.4 shows the performance of the XGBoost model after applying custom hyperparameter optimization as discussed from the previous chapter. The updated model tracks the actual load more closely across all days, showing improved responsiveness to changes in demand. As a result, the MAE was reduced to 2039.53 and the MAPE dropped to 2.33%. This indicates significant improvement in the forecast accuracy. Moreover, this comparison shows the importance of model tuning when applying ensemble-based methods like XGBoost to time series forecasting tasks. The final forecasting results of the XGBoost algorithm are tabulated in Table 6.3 below:

Table 6.3: XGBoost Model Forecast Performance.

| Model   | MAE           | MAPE  |
|---------|---------------|-------|
| XGBoost | 2039.53 (MWh) | 2.33% |

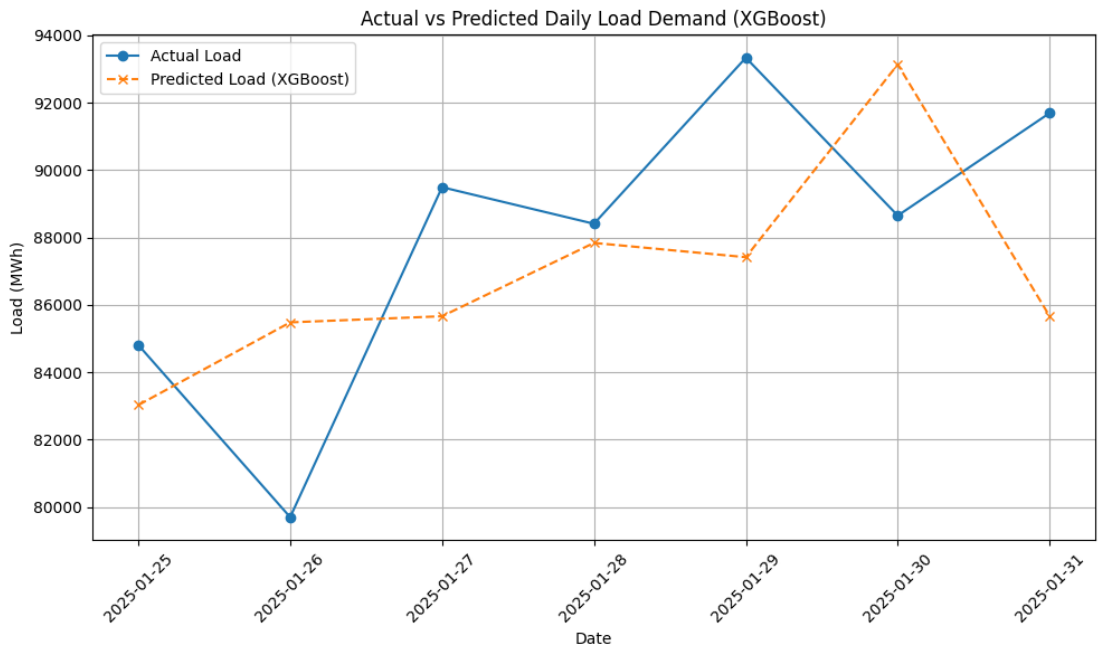

Figure 6.3: XGBoost Forecast vs Actual Load without Customized Hyperparameters.

The optimized XGBoost model, as shown in Fig. 6.4, produced better forecasting results across the entire forecast window. Unlike the earlier version, the model better learned the overall trends and daily variations with minimal deviation from the actual values. This concludes that the improved model was better altered to the load demand's behavior and was able to make more precise adjustments based on recent load and weather patterns as discussed earlier.

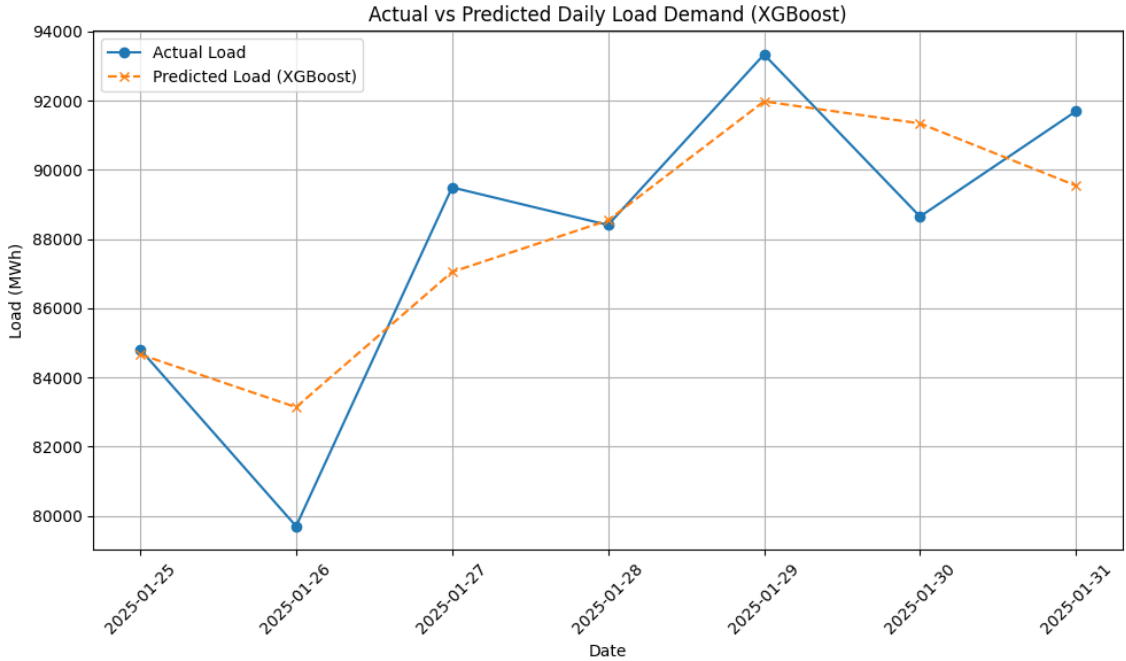

Figure 6.4: Optimized XGBoost Forecast vs Actual Load with Custom Hyperparameters.

### 6.6 LightGBM Model Results

The results of the LightGBM model are presented in Figure 6.5, where the predicted daily load values are plotted alongside the actual observed data. This model delivered strong forecasting performance, achieving a mean absolute error (MAE) of 1708.22 and a mean absolute percentage error (MAPE) of 1.95%. Moreover, the forecast line follows the trend of the actual demand across all seven days, with minimal deviation during both rising and falling intervals.

The improved accuracy of LightGBM can be attributed to both its architecture and the use of customized hyperparameters as discussed earlier. Like the customized

XGBoost model, this implementation was also configured with task-specific settings that composed learning depth, tree complexity, and regularization. However, LightGBM inherently uses a histogram-based approach and processes leaf-wise rather than level-wise, which tends to result in faster training and more efficient splits when handling structured data like time series. The final forecasting results of the LightGBM algorithm are tabulated in Table 6.4 below:

Table 6.4: LightGBM Model Forecast Performance.

| Model | MAE | MAPE |
| --- | --- | --- |
| LightGBM | 1708.22 (MWh) | 1.95% |

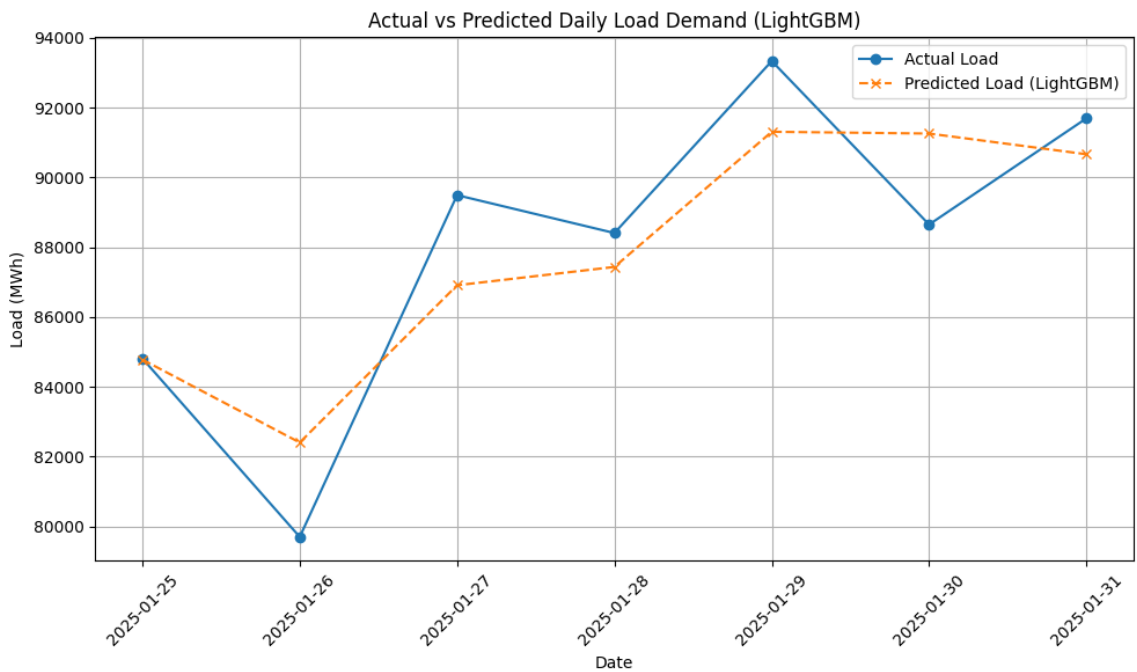

Figure 6.5: Optimized LightGBM Forecast vs Actual Load using Custom Hyperparameters.

For comparison, the initial version of LightGBM without the tuned parameters achieved a MAPE of 2.18% and MAE of 1984.5, which, while still strong, was slightly outperformed by the optimized configuration. The comparison concludes that while LightGBM performs well out-of-the-box, optimized tuning further improves its precision and forecasting abilities. In conclusion, the LightGBM model worked very well in this study and can be considered as one of the better GBDT models in short-term load forecasting.

*6.7 LSTM Model Results*

The results of the LSTM model are illustrated in Figure 6.6, where the predicted daily load is plotted alongside the actual demand for the same 7-day period. The model achieved a mean absolute error (MAE) of 1535.71 and a mean absolute percentage error (MAPE) of 1.74% tabulated in Table 6.5 in the following page.

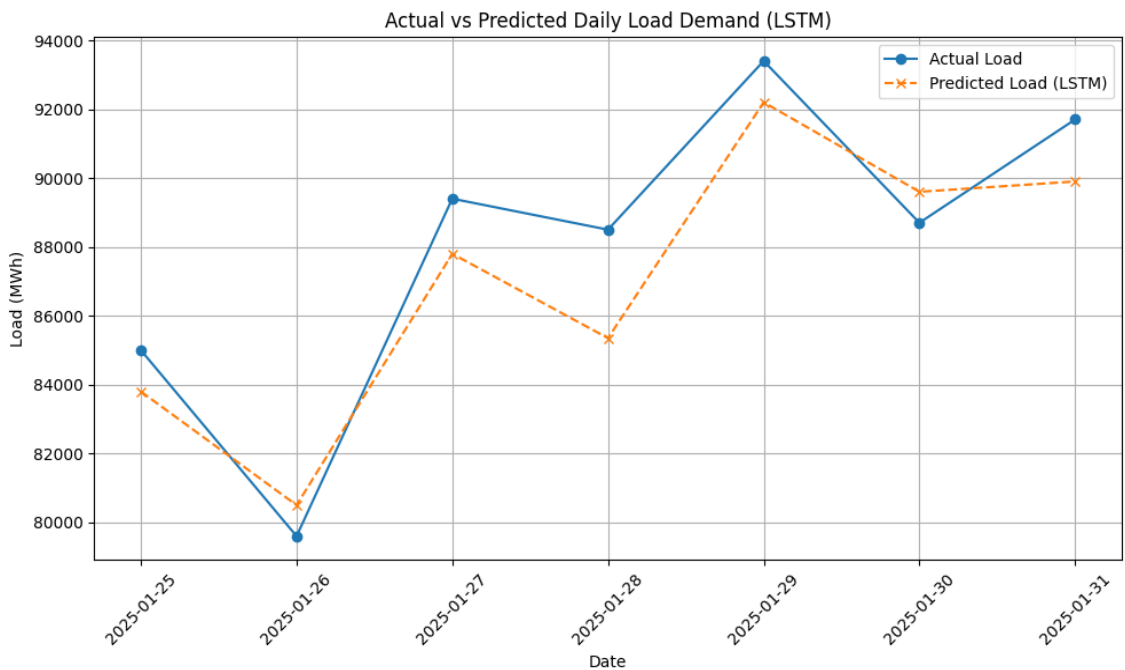

Figure 6.6: LSTM Model Forecast vs Actual Load Results.

As shown in Fig 6.5 and summarized in Table 6.4, the LSTM model achieved the highest numerical accuracy among all models tested. However, there are certain caveats worth discussing regarding the LSTM approach.

Table 6.5: LSTM Model Forecast Performance.

| Model | MAE | MAPE |
|-------|-----|------|
| LSTM | 1535.71 (MWh) | 1.74% |

While the LSTM model produced great accuracy, it is important to consider the computational cost associated with its training and tuning of such model [44, 45]. Compared to the LightGBM model, which achieved a slightly higher MAPE of 1.95 percent and MAE of 1708.22, the LSTM model requires significantly more time and computing resources due to its sequential structure and reliance on GPU/CPU cycles for backpropagation through time. LightGBM, in contrast, computes much faster and still offers respectable performance.

In conclusion, the LSTM model architecture designed in this study proved to be an accurate forecasting candidate for short-term load forecasting due to its ability to learn temporal dependencies from sequential data. Although the model's performance varied slightly depending on the configuration, it steadily produced accurate forecasting results and followed the shape of actual load behavior as shown in Fig. 6.6. Overall, the LSTM approach confirmed that memory-based models can effectively forecast daily electricity demand when provided with well-prepared input sequences and an appropriate training setup which was accomplished in this study.

*6.8 GRU Model Results*

The results of the GRU model are shown in Figure 6.7. The GRU forecast follows the overall trend of the actual load well, including the main turning points and shifts in the load direction. The GRU model achieved a mean absolute error (MAE) of 2178.57 and a mean absolute percentage error (MAPE) of 2.53% after optimization.

Table 6.6: GRU Model Forecast Performance.

| Model | MAE | MAPE |
|---|---|---|
| GRU | 2178.57 (MWh) | 2.53% |

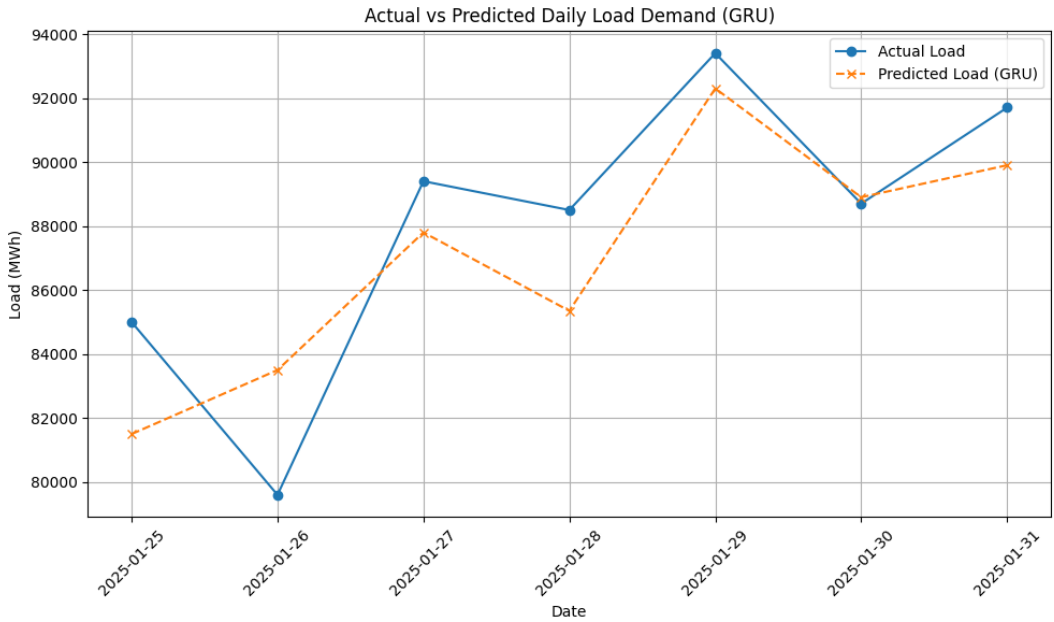

Figure 6.7: GRU Model Forecast vs Actual Load Results.

Compared to the LSTM model, which achieved a lower MAPE of 1.74%, the GRU model was slightly less accurate. While both models share similar recurrent structures and were trained using the same features and time window, the GRU's overall simpler architecture may have contributed to the small performance gap.

# CHAPTER 7

## DISCUSSION & FUTURE PROSPECTS

### *7.1 Summary of Results*

The load forecasting performance of each model evaluated and implemented in this study is summarized in Table 7.1.

Table 7.1: Load Forecasting Performance of All Models.

| Model | MAE (MWh) | MAPE (%) |
|---|---|---|
| Baseline ARIMA | 3878.2 | 4.03% |
| GRU | 2178.57 | 2.53% |
| XGBoost | 2039.53 | 2.33% |
| Hybrid ARIMA-SVM | 1857.14 | 2.09% |
| LightGBM | 1708.22 | 1.95% |
| LSTM | 1535.71 | 1.74% |

The final results of this study indicate that all models outperform the baseline ARIMA method, which shows the highest error rates with an MAE of 3878.2 MWh and a MAPE of 4.03%. Among the RNNs, LSTM achieved the most desirable results in this work, with the lowest errors across both metrics. However, LSTM models, while producing lower error, are generally more computationally intensive and require greater resources compared to the other methods. LightGBM and XGBoost show strong forecasting performance, with relatively low MAE and MAPE values, while GRU performs moderately well, slightly trailing behind LSTM and LightGBM. Although the LSTM model gives the better overall results, LightGBM with customized hyperparameters can also serve as an effective alternative approach, one that is more computationally efficient with good accuracy in short-term load forecasting.

## 7.3 Future Research

From a modeling perspective, research can explore more hybrid frameworks that combine sequence models (i.e., LSTM, GRU) with decision-tree ensembles to influence both temporal dependencies and non-linear feature interactions. While this study implemented a hybrid ARIMA-SVM model, developing additional hybrid models based on different frameworks can be worth the research. For instance, one of the primary challenges of a hybrid model comes from the mismatch between the sequential nature of recurrent models and the tabular structure expected by tree-based algorithms. Moreover, in short-term load forecasting, incremental retraining methods can also be further studied for maintaining model performance as new data becomes available. Regularization techniques, dropout scheduling, and ensemble averaging should be researched to address overfitting risks. Future research in short-term load forecasting can also benefit from more reliable data acquisition by the balancing authorities. Better data reduces the overall preprocessing time and as such, researchers can focus more on the forecasting algorithms rather than making the data reliable. Lastly, external data pipelines such as weather APIs can be encoded using temporal indicator matrices, or embedded vectors. This can generate more accurate forecasting results since it is relying on real-time data, as past data can't always reference to future or present events. These future research directions are offered based on the work that was done in this research and the challenges encountered in the modeling and forecasting process.

CHAPTER 8

CONCLUSION

In this study, several load forecasting techniques were implemented and compared in order to evaluate their effectiveness in predicting short-term power load demand for the State of Arizona. The models included a traditional time series method, ARIMA, hybrid ARIMA-SVM, and four machine learning-based frameworks: XGBoost, LightGBM, LSTM, and GRU. The forecasting task focused on daily load demand, and the goal was to assess how well each forecasting method could generalize and predict unseen values using historical load and weather data. The study incorporated feature analysis using PCC, which helped classify the impact of exogenous variables with the load demand behavior. The baseline ARIMA model achieved a MAPE of 4.03%. However, the hybrid ARIMA-SVM model improved upon the baseline ARIMA and achieved a MAPE of 2.09%. Moreover, LSTM delivered the most accurate forecasting performance in this study with a MAPE of 1.74%, followed by LightGBM at 1.95% in this research. GRU and XGBoost had slightly higher errors of 2.53% and 2.33% respectively. The findings from this study suggest that both memory-based and tree-based models perform well in short-term load forecasting when configured with the appropriate hyperparameters and trained on relevant historical and external features. In closing, while all models delivered solid forecasting performance, there is no single machine learning approach from this study that clearly outperformed the rest in every aspect. Each model brought its own strengths to the task, and their success depended on how they were configured, trained, and on the nature of the input data.

REFERENCES

[1] U.S. DOE, "FOTW #1365, October 21, 2024: U.S. Net Generation of Electricity Relied on Record Use of Renewables while Coal Use Dropped to a Record Low in 2023," *Office of Energy Efficiency & Renewable Energy*, Oct. 21, 2024.

[2] I. S. Jahan, V. Snasel, and S. Misak, "Intelligent Systems for Power Load Forecasting: a Study Review," *Energies*, vol. 13, no. 22, p. 6105, Nov. 2020, doi: https://doi.org/10.3390/en13226105.

[3] Seyed Azad Nabavi, S. Mohammadi, Naser Hossein Motlagh, Sasu Tarkoma, and P. Geyer, "Deep Learning Modeling in Electricity Load forecasting: Improved Accuracy by Combining DWT and LSTM," *Energy Reports*, vol. 12, no. 2, pp. 2873–2900, Dec. 2024, doi: https://doi.org/10.1016/j.egyr.2024.08.070.

[4] J. Zheng, C. Xu, Z. Zhang, and X. Li, "Electric Load Forecasting in Smart Grids Using Long-Short-Term-Memory Based Recurrent Neural Network," *2017 51st Annual Conference on Information Sciences and Systems (CISS)*, vol. i, no. 1, Mar. 2017, doi: https://doi.org/10.1109/ciss.2017.7926112.

[5] M. Gao, S. Zhou, W. Gu, Z. Wu, H. Liu, and A. Zhou, "A General Framework for Load Forecasting Based on Pre-trained Large Language Model," *Arxiv.org*, 2018. https://arxiv.org/html/2406.11336v2.

[6] D. Ortiz-Arroyo, M. K. Skov, and Q. Huynh, "Accurate Electricity Load Forecasting with Artificial Neural Networks," *Computational Intelligence for Modelling, Control and Automation,* 2005. pp. 94–99, Dec. 2005, doi: https://doi.org/10.1109/CIMCA.2005.1631248.

[7] T. Xue, U. Karaagac, I. Kocar, M. B. Vavdareh, and M. Ghafouri, "Machine Learning Basics and Potential Applications in Power Systems," *2023 International Conference on Electrical, Communication and Computer Engineering (ICECCE)*, pp. 1–7, Dec. 2023, doi: https://doi.org/10.1109/icecce61019.2023.10441935.

[8] P. Kaledio, "Machine Learning Applications in Electric Power Systems: Enhancing Efficiency, Reliability, and Sustainability," *SSRN Electronic Journal*, Jan. 2024, doi: https://doi.org/10.2139/ssrn.4716389.

[9] E. A. Feinberg and D. Genethliou, "Load Forecasting in Applied Mathematics for Power Systems," *Power Electronics and Power Systems*, pp. 269–285, Aug. 2019, doi: https://doi.org/10.1007/0-387-23471-3_12.

[10] G. Nalcaci, A. Özmen, and G. W. Weber, "Long-term Load forecasting: Models Based on MARS, ANN and LR Methods," *Central European Journal of Operations Research*, vol. 27, no. 4, pp. 1033–1049, Mar. 2018, doi: https://doi.org/10.1007/s10100-018-0531-1.

[11] M. A. Hammad, B. Jereb, B. Rosi, and D. Dragan, "Methods and Models for Electric Load Forecasting: A Comprehensive Review," *Logistics & Sustainable Transport*, vol. 11, no. 1, pp. 51–76, Feb. 2020, doi: https://doi.org/10.2478/jlst-2020-0004.

[12] N. Ahmad, Y. Ghadi, M. Adnan, and M. Ali, "Load Forecasting Techniques for Power System: Research Challenges and Survey," *IEEE Xplore*, pp. 1–15, Jul. 2022, doi: https://doi.org/10.1109/access.2022.3187839.

[13] M. G. Pinheiro, S. C. Madeira, and A. P. Francisco, "Short-term Electricity Load forecasting—A Systematic Approach from System Level to Secondary Substations," *Applied Energy*, vol. 332, pp. 120493–120493, Feb. 2023, doi: https://doi.org/10.1016/j.apenergy.2022.120493.

[14] R. K. Agrawal, F. Muchahary, and M. M. Tripathi, "Long Term Load Forecasting with Hourly Predictions Based on long-short-term-memory Networks," *IEEE Xplore*, Feb. 18, 2018. https://ieeexplore.ieee.org/stamp/stamp.jsp?tp=&arnumber=8312088.

[15] S. Chen, R. Lin, and W. Zeng, "Short-Term Load Forecasting Method Based on ARIMA and LSTM," *2022 IEEE 22nd International Conference on Communication Technology (ICCT)*, pp. 1913–1917, Mar. 2023, doi: https://doi.org/10.1109/icct56141.2022.10073051.

[16] R. J. Hyndman and G. Athanasopoulos, "Forecasting: Principles and Practice," *Otexts.com*, 2018. https://otexts.com/fpp2/.

[17] G. E. P. Box and G. M. Jenkins, "Time Series Analysis: Forecasting and Control," 1970.

[18] A. Al Mamnun, MD. Sohel, N. Mohammad, S. Haque, and D. Roy Dipta, "A Comprehensive Review of the Load Forecasting Techniques Using Single and Hybrid Predictive Models," Jun. 26, 2020. https://ieeexplore.ieee.org/stamp/stamp.jsp?arnumber=9144528.

[19] M. A. A. Amin and Md. A. Hoque, "Comparison of ARIMA and SVM for Short-term Load Forecasting," *IEEE Xplore*, Oct. 21, 2019. https://ieeexplore.ieee.org/stamp/stamp.jsp?tp=&arnumber=8877077.

[20]   N. Rani, S. K. Aggarwal, and S. Kumar, "Short-Term Load Foresting Using Combination of Linear and Non-linear Models," *IEEE Access*, vol. 1, no. 1, pp. 1–1, Apr. 2024, doi: https://doi.org/10.1109/access.2024.3392592.

[21]   H. Kaur and S. Ahuja, "SARIMA Modelling for Forecasting the Electricity Consumption of a Health Care Building," *International Journal of Innovative Technology and Exploring Engineering*, vol. 8, no. 12, pp. 2795–2799, Oct. 2019, doi: https://doi.org/10.35940/ijitee.l2575.1081219.

[22]   E. Chodakowska, J. Nazarko, and Ł. Nazarko, "ARIMA Models in Electrical Load Forecasting and Their Robustness to Noise," *Energies*, vol. 14, no. 23, p. 7952, Nov. 2021, doi: https://doi.org/10.3390/en14237952.

[23]   N. Tang and D.-J. Zhang, "Application of a Load Forecasting Model Based on Improved Grey Neural Network in the Smart Grid," *Energy Procedia*, vol. 12, pp. 180–184, 2011, doi: https://doi.org/10.1016/j.egypro.2011.10.025.

[24]   N. Amral, C. S. Ozveren, and D. King, "Short Term Load Forecasting Using Multiple Linear Regression," *IEEE Xplore*, Oct. 04, 2007. https://ieeexplore.ieee.org/document/4469121.

[25]   C. Tarmanini, N. Sarma, C. Gezegin, and O. Ozgonenel, "Short Term Load Forecasting Based on ARIMA and ANN Approaches," *Energy Reports*, vol. 9, no. 3, pp. 550–557, May 2023, doi: https://doi.org/10.1016/j.egyr.2023.01.060.

[26]   B. Emre Türkay and D. Demren, "Electrical load forecasting using support vector machines," *7th International Conference on Electrical and Electronics Engineering (ELECO)*, vol. 1, no. 1, Dec. 2011, doi: https://ieeexplore.ieee.org/document/6140142.

[27]   X. Li, C. Sun, and D. Gong, "Application of Support Vector Machine and Similar Day Method for Load Forecasting," *Springer Nature Link*, vol. 1, no. 1, pp. 602–609, 2005, doi: https://doi.org/10.1007/11539117_85.

[28]   H. S. Hippert, C. E. Pedreira, and R. C. Souza, "Neural Networks for short-term Load forecasting: A Review and Evaluation," *IEEE Transactions on Power Systems*, vol. 16, no. 1, pp. 44–55, 2001, doi: https://doi.org/10.1109/59.910780.

[29]   A. M. Hemeida *et al.*, "Nature-inspired Algorithms for feed-forward Neural Network classifiers: a Survey of One Decade of Research," *Ain Shams Engineering Journal*, vol. 11, no. 3, pp. 659–675, Sep. 2020, doi: https://doi.org/10.1016/j.asej.2020.01.007.

[30]  K. Wang, J. Zhang, X. Li, and Y. Zhang, "Long-Term Power Load Forecasting Using LSTM-Informer with Ensemble Learning," *Electronics*, vol. 12, no. 10, p. 2175, May 2023, doi: https://doi.org/10.3390/electronics12102175.

[31]  X. Yu, Z. Xu, X. Zhou, J. Zheng, Y. Xia, and L. Lin, "Load Forecasting Based on Smart Meter Data and Gradient Boosting Decision Tree," *2019 Chinese Automation Congress (CAC)*, vol. 1, no. 1, Feb. 2020, doi: https://doi.org/10.1109/cac48633.2019.8996810.

[32]  P. Nie, M. Roccotelli, M. P. Fanti, Z. Ming, and Z. Li, "Prediction of Home Energy Consumption Based on Gradient Boosting Regression Tree," *Energy Reports*, vol. 7, pp. 1246–1255, Nov. 2021, doi: https://doi.org/10.1016/j.egyr.2021.02.006.

[33]  L. Di Persio and N. Fraccarolo, "Energy Consumption Forecasts by Gradient Boosting Regression Trees," *Mathematics*, vol. 11, no. 5, p. 1068, Feb. 2023, doi: https://doi.org/10.3390/math11051068.

[34]  S. Muzaffar and A. Afshari, "Short-Term Load Forecasts Using LSTM Networks," *Energy Procedia*, vol. 158, pp. 2922–2927, Feb. 2019, doi: https://doi.org/10.1016/j.egypro.2019.01.952.

[35]  M. Abumohsen, A. Y. Owda, and M. Owda, "Electrical Load Forecasting Using LSTM, GRU, and RNN Algorithms," *Energies*, vol. 16, no. 5, p. 2283, Feb. 2023, doi: https://doi.org/10.3390/en16052283.

[36]  R. Bareth, A. Yadav, S. Gupta, and M. Pazoki, "Daily Average Load Demand Forecasting Using LSTM Model Based on Historical Load Trends," *IET Generation, Transmission & Distribution*, vol. 18, no. 5, pp. 952–962, Feb. 2024, doi: https://doi.org/10.1049/gtd2.13132.

[37]  Z. Chen *et al.*, "Load Forecasting Based on LSTM Neural Network and Applicable to Loads of 'Replacement of Coal with Electricity,'" *Journal of Electrical Engineering & Technology*, vol. 16, no. 5, pp. 2333–2342, Apr. 2021, doi: https://doi.org/10.1007/s42835-021-00768-8.

[38]  W. Kong, Z. Y. Dong, Y. Jia, D. J. Hill, Y. Xu, and Y. Zhang, "Short-Term Residential Load Forecasting Based on LSTM Recurrent Neural Network," *IEEE Transactions on Smart Grid*, vol. 10, no. 1, pp. 841–851, Oct. 2017, doi: https://doi.org/10.1109/tsg.2017.2753802.

[39]  M. M. Hossain and H. Mahmood, "Short-Term Load Forecasting Using an LSTM Neural Network," *2020 IEEE Power and Energy Conference at Illinois (PECI)*, Feb. 2020, doi: https://doi.org/10.1109/peci48348.2020.9064654.

[40]  C. Olah, "Understanding LSTM Networks," *Colah's Blog*, Aug. 27, 2015. https://colah.github.io/posts/2015-08-Understanding-LSTMs/.

[41]  H. Li, Y. Weng, and H. Tong, "Heterogeneous Transfer Learning on Power Systems: A Merged Multi-modal Gaussian Graphical Model," *2021 IEEE International Conference on Data Mining (ICDM)*, Nov. 2020, doi: https://doi.org/10.1109/icdm50108.2020.00130.

[42]  N. Bashiri Behmiri, C. Fezzi, and F. Ravazzolo, "Incorporating Air Temperature into mid-term Electricity Load Forecasting Models Using time-series Regressions and Neural Networks," *Energy*, vol. 278, p. 127831, Sep. 2023, doi: https://doi.org/10.1016/j.energy.2023.127831.

[43]  S. Khatoon, Ibraheem, A. K. Singh, and Priti, "Effects of Various Factors on Electric Load forecasting: an Overview," *2014 6th IEEE Power India International Conference (PIICON)*, Jun. 2015, doi: https://doi.org/10.1109/poweri.2014.7117763.

[44]  S. Merity, N. Shirish Keskar, and R. Socher, "Regularizing and Optimizing LSTM Language Models," *arXiv*, vol. 1, no. 1, Aug. 2017, Available: https://arxiv.org/pdf/1708.02182

[45]  K. Khalil, O. Eldash, A. Kumar, and M. Bayoumi, "Economic LSTM Approach for Recurrent Neural Networks," *IEEE Transactions on Circuits & Systems II Express Briefs*, vol. 66, no. 11, pp. 1885–1889, Jun. 2019, doi: https://doi.org/10.1109/tcsii.2019.2924663.

[46]  K. Khalil, O. Eldash, A. Kumar, and M. Bayoumi, "Economic LSTM Approach for Recurrent Neural Networks," *IEEE Transactions on Circuits & Systems II Express Briefs*, vol. 66, no. 11, pp. 1885–1889, Jun. 2019, doi: https://doi.org/10.1109/tcsii.2019.2924663.

[47]  J. Xie, Inalvis Alvarez-Fernandez, and W. Sun, "A Review of Machine Learning Applications in Power System Resilience," *IEEE Xplore*, vol. 1, no. 1, Aug. 2020, doi: https://doi.org/10.1109/pesgm41954.2020.9282137.

[48]  H. Nie, G. Liu, X. Liu, and Y. Wang, "Hybrid of ARIMA and SVMs for Short-Term Load Forecasting," *Energy Procedia*, vol. 16, pp. 1455–1460, 2012, doi: https://doi.org/10.1016/j.egypro.2012.01.229.

[49]  S. Karthika, V. Margaret, and K. Balaraman, "Hybrid Short Term Load Forecasting Using ARIMA-SVM," *IEEE Xplore*, Apr. 01, 2017. https://ieeexplore.ieee.org/document/8245060.

[50] A. Sinha, R. Tayal, A. Vyas, P. Pandey, and O. P. Vyas, "Forecasting Electricity Load with Hybrid Scalable Model Based on Stacked Non Linear Residual Approach," *Frontiers in Energy Research*, vol. 9, no. 1, Nov. 2021, doi: https://doi.org/10.3389/fenrg.2021.720406.

[51] A. A. El Desouky and M. M. El Kateb, "Hybrid Adaptive Techniques for electric-load Forecast Using ANN and ARIMA," *IEE Proceedings - Generation, Transmission and Distribution*, vol. 147, no. 4, p. 213, 2010, doi: https://doi.org/10.1049/ip-gtd:20000521.

APPENDIX A

ACRONYMS AND ABBREVIATIONS

ACF: Autocorrelation Function

ANN: Artificial Neural Network

APS: Arizona Public Service

AR: Autoregressive

ARIMA: Autoregressive Integrated Moving Average

ARMA: Autoregressive Moving Average

BA: Balancing Authority

EIA: U.S. Energy Information Administration

GBDT: Gradient Boosting Decision Tree(s)

GRU: Gradient Recurring Unit

kW: Kilowatt

kWh: Kilowatt-Hour

LF: Load Forecasting

LightGBM: Light Gradient Boosting

LSTM: Long Short-Term Memory

LTLF: Long-Term Load Forecasting

MA: Moving Average

MAE: Mean Absolute Error

MAPE: Mean Absolute Percentage ERROR

ML: Machine Learning

MLR: Multiple Linear Regression

MTLF: Medium-Term Load Forecasting

MW: Megawatt

MWh: Megawatt-Hour

NOAA: National Oceanic and Atmospheric Administration

PACF: Partial Auto-Correlation Function

PCC: Pearson Correlation Coefficient

RMSE: Root Mean Square Error

RNN: Recurrent Neural Network

SARIMA: Seasonal Autoregressive Integrated Moving Average

STLF: Short-Term Load Forecasting

SVR: Support Vector Regression

SVM: Support Vector Machine(s)

TAVG: Temperature Average

TMAX: Temperature Maximum

TMIN: Temperature Minimum

VSTLF: Very Short-Term Load Forecasting

XGBOOST: eXtreme Gradient Boosting